\documentclass[a4paper,aps,superscriptaddress,floatfix,nofootinbib,twocolumn,longbibliography]{revtex4-1}

\usepackage{amsmath, amsthm, amssymb} 
\usepackage{algpseudocode} 
\usepackage{comment} 
\usepackage{graphicx}
\usepackage{dcolumn}
\usepackage{bm}
\usepackage{xcolor}
\usepackage{hyperref}
\usepackage[mathlines]{lineno}
\usepackage{float}

\newcommand{\change}[1]{{#1}}
\newcommand{\newchange}[1]{{#1}}

\theoremstyle{definition}

\begin{document}

\title{Epidemic spreading in group-structured populations}

\author{Siddharth Patwardhan}
\thanks{These authors contributed equally to this work.}
\affiliation{Center for Complex Networks and Systems Research, Luddy School of Informatics, Computing, and Engineering, Indiana University, Bloomington, Indiana 47408, USA}

\author{Varun K. Rao}
\thanks{These authors contributed equally to this work.}
\affiliation{Center for Complex Networks and Systems Research, Luddy School of Informatics, Computing, and Engineering, Indiana University, Bloomington, Indiana 47408, USA}
\affiliation{School of Public Health, Indiana University, Bloomington, Indiana 47408, USA}

\author{Santo Fortunato}
\affiliation{Center for Complex Networks and Systems Research, Luddy School of Informatics, Computing, and Engineering, Indiana University, Bloomington, Indiana 47408, USA}

\author{Filippo Radicchi}
\affiliation{Center for Complex Networks and Systems Research, Luddy School
  of Informatics, Computing, and Engineering, Indiana University, Bloomington,
  Indiana 47408, USA}
\email{filiradi@indiana.edu}

\begin{abstract}
Individuals involved in common group activities/settings -- e.g., college students that are enrolled in the same class and/or live in the same dorm -- are exposed to recurrent contacts of physical proximity. These contacts are known to mediate the spread of an infectious disease, however, it is not obvious how the properties of the spreading process are determined by the structure of and the interrelation among the group settings that are at the root of those recurrent interactions.
Here, we show that reshaping the organization of groups within a population can be used as an effective strategy to decrease the severity of an epidemic. Specifically, we show that when group structures are sufficiently correlated -- e.g., the likelihood for two students living in the same dorm to attend the same class is sufficiently high -- outbreaks are longer but milder than for uncorrelated group structures. Also, we show that the effectiveness of interventions for disease containment increases as the correlation among group structures increases. We demonstrate the practical relevance of our findings by taking advantage of data about housing and attendance of students at the Indiana University campus in Bloomington. By appropriately optimizing the assignment of students to dorms based on their enrollment, we are able to observe a two- to five-fold reduction in the severity of simulated epidemic processes.
\end{abstract}

\maketitle


\section{Introduction}

From the Black Plague in the 13th century to the recent COVID-19 pandemic, epidemics have always 
represented significant threats for
 humanity~\cite{wade2020black}. 
 Network science has been at the core of many advances made in epidemic modeling, given that the fate of an epidemic process is fundamentally determined by the structure of the social network where spreading occurs~\cite{Pastor-Satorras2015,vespignani2020modelling,liu2018measurability,soriano2018spreading}. Notably, 
 network structural properties
 play non-trivial roles in shaping not only the dynamics of spreading, but also 
  mitigation strategies~\cite{Pastor-Satorras2015,masuda2009immunization,pastor2002immunization,chen2008finding,pastor2003epidemics,osat2023embedding}.

Contacts of physical proximity that mediate spreading among individuals often occur in group activities/settings~\cite{granell2018epidemic, gomez2018critical,soriano2018spreading}.
For example in a college, two students can get in contact because they are enrolled in the same class and/or live in the same dorm. Similarly,  physical proximity between school-going children happen in classroom/family settings. Realistic models of epidemic spreading are informed by data accounting for the existence of multiple types of group interactions that individuals are exposed to~\cite{broeck2011gleamviz}.
This aspect is also accounted for in some recent theoretical meta-population models of epidemic spreading~\cite{Granell2013,Gomez2013,granell2014competing,soriano2018spreading,gomez2018critical,granell2018epidemic}.

A common feature of the above-mentioned studies is considering the exposure of individuals to multiple types of group interactions as an input rather than a free\change{, or tunable,} parameter of the spreading model.
\newchange{For example, Granell and Mucha assume that the groups 
of a population are given and then analytically study how the threshold value of an epidemic spreading in the group-structured population depends on the mobility of individuals among groups~\cite{granell2018epidemic}.}
In this paper, we change perspective by focusing our attention on the effects that group structure and interrelation among groups have on the properties of a Susceptible-Infected-Recovered (SIR) epidemic process. To this end, we represent interaction patterns among individuals of a population as edge-colored graphs~\cite{ramsey1987problem} with block/community structure~\cite{holland1983stochastic}. Each color or layer in these graphs represents a specific social setting where two individuals may get in physical proximity, e.g., a classroom or a dorm in the case of college students. Blocks or communities are instead used to model group interactions in the layers. In the example of a population of college students, blocks represent specific classes attended by students or specific dorms where students reside in. The framework allows us to control for the strength of the block structure of the layers, e.g., the likelihood that two students in the same class/dorm have a proximity contact, as well as for the similarity of groups across layers, e.g., the likelihood that two students live in the same dorm and attend the same class. We study the importance of these two factors in determining the fate of epidemic processes.

A study that is immediately related to ours is the one  by Fan \textit{et al.}\cite{Fan2019}. They consider epidemic spreading in edge-colored graphs that are embedded in geometric space.  They numerically estimate the epidemic threshold for graphs with variable levels of correlation between the layers' embeddings, and find that such a quantity grows as geometric correlation increases. In other words, the severity of an epidemic is reduced in a network with geometrically correlated layers compared to a network whose layers are not correlated.

As there exists a tight analogy between geometric embedding and community structure~\cite{faqeeh2018characterizing}, the work by Fan \textit{et al.} 
directly relates to ours. 
In our work, however,
\change{we go well beyond the state of the art by} 
expanding the analysis by Fan \textit{et al.} in two main, fundamental respects. First, we provide a full characterization of the importance of group strength and correlation in epidemic spreading by monitoring not only static quantities such as the size of the outbreak and the epidemic threshold, but also dynamical quantities such as the duration and intensity of the epidemic.
\change{Looking at dynamical metrics of epidemic severity turns out to be essential for a full understanding of how topological correlations affect spreading dynamics.}
In particular, there are cases where changes in group correlation do not lead to \change{any}
variations in outbreak size and \change{epidemic threshold}, 
but to apparent variations of both duration and intensity.
In addition, we consider the role of correlation in  interventions for disease containment, \change{an aspect that is not considered by Fan {\it et al.}, but that actually displays a marked dependence on the topological correlations that characterize a network.} We find that herd immunity in a network with correlated group structure can be achieved by immunizing a fraction of individuals that is even five-fold smaller than in the case of a network with non-correlated group structure. 
\change{Second and more important, Fan et al. do not apply their approach to the analysis of real-world systems. The method is in fact limited by a series of challenges in practical applications. The approach requires knowledge of the network topology of the layers to perform the embedding in the hyperbolic space, but detailed network data are in many cases not readily available. Also, once the embedding is performed, manipulating the network to reduce the severity of a disease outbreak is not an intuitive task as it requires changing coordinates of nodes in the hyperbolic space.  Our approach instead does not require knowledge of the network of contacts. We simply use information about groups' composition. This information is generally available in real systems. Further, interventions on the network are easy to formulate/implement as they require only changing the group assignments of nodes. In this paper, we actually demonstrate the utility of our proposed framework in the analysis of }
a real-world scenario of data concerning enrollment and housing of college students at the Indiana University campus in Bloomington. We show that a simple strategy of re-assigning students to housing facilities  can (i) dramatically reduce the severity of an epidemic outbreak, both in terms of intensity and duration, as well as (ii) greatly enhance the effectiveness of interventions for disease containment.

\section{Results}

\subsection{Synthetic graphs}

\begin{figure}[!htb]
    \centering
    \includegraphics[width=0.48\textwidth]{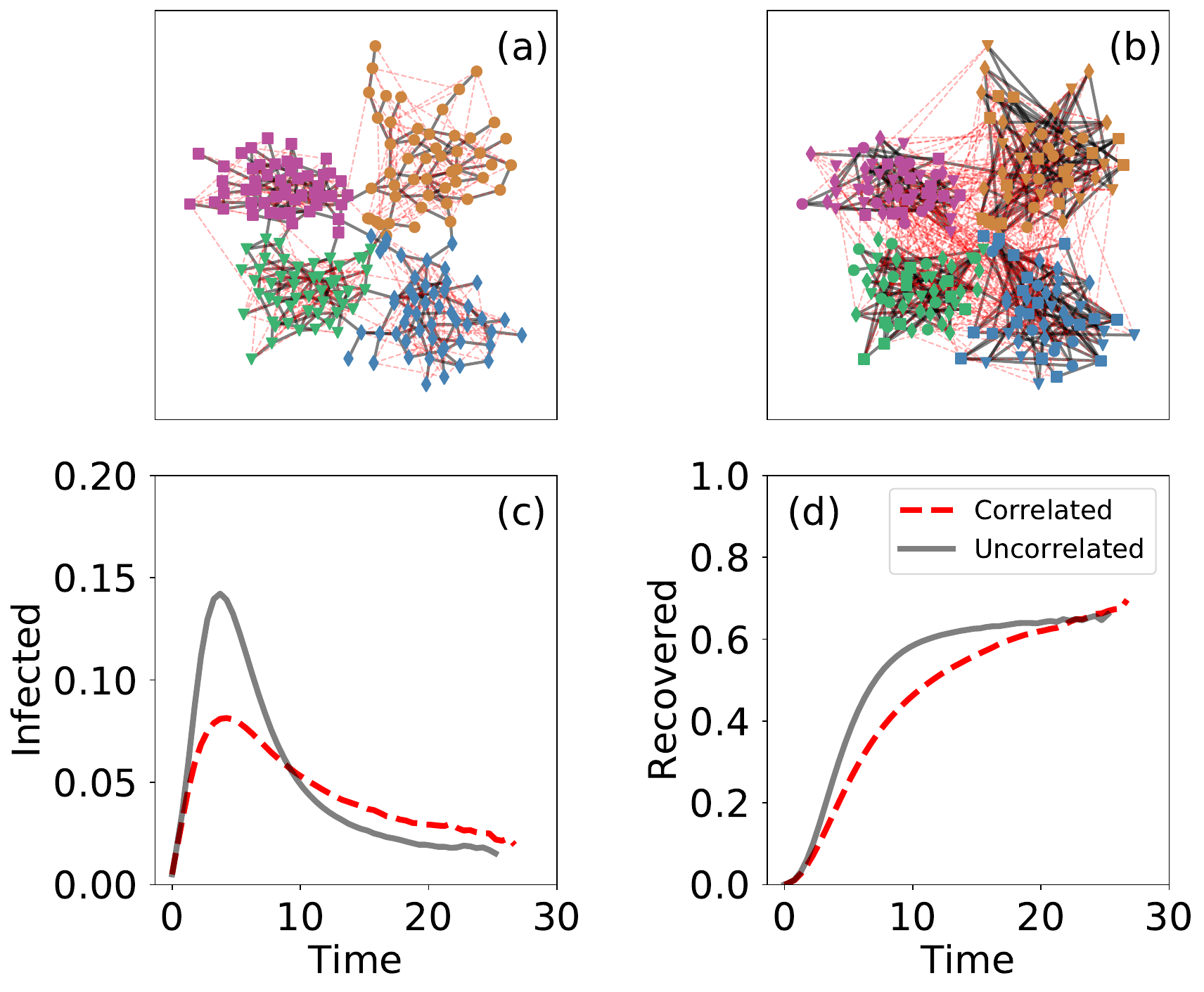}
    \caption{{\bf Modeling framework and metrics of epidemic severity.} (a) As a basic illustration of the modeling framework, we consider an edge-colored graph composed of  $N=200$ nodes and two layers of interactions. Nodes are organized in four groups of size $q^{(\ell)} = 50$ for both layer $\ell=1$ and $\ell=2$. Colors of the nodes (i.e., orange, purple, blue, green) represent the community structure in layer $\ell=1$; shapes (i.e., circle, square, triangle, diamond) indicate the community memberships in layer $\ell=2$. Here, the community structures of the layers are maximally correlated (NMI = $1.0$, with NMI standing for normalized mutual information), meaning that there is a one-to-one map from shapes to colors. Connections among pairs of nodes are created such that the average degree is $\langle k^{(\ell)} \rangle = 3$; a fraction $\mu = 0.025$ of these edges connects a node to other nodes outside its own community. Using these parameters, we generate edges in layer $\ell=1$ (full gray) and in layer $\ell=2$ (dashed red). (b) Same as in (a), but for uncorrelated community structure (NMI = $0.0$), i.e., colors and shapes are assigned to nodes randomly. (c) We run \newchange{$V=1,000,000$} simulations of the SIR model on top of the graphs of panels (a) and (b) by setting the spreading rate $\beta = 0.4$\change{, i.e., setting the basic reproduction number $R_0 = \beta (\langle k^{(1)} \rangle+\langle k^{(2)} \rangle) = 2.4$}. We display the fraction of infected nodes in the population as a function of time. We consider bins of size $0.05$ and only report the average values over \newchange{at least $100$} surviving runs. (d) Same as in (c), but here we are displaying the fraction of recovered nodes in the population as a function of time.}
    \label{fig:0}
\end{figure}

In Fig.~\ref{fig:0}, we provide an example of our modeling framework, see Methods for details. Continuous-time Susceptible-Infected-Recovered (SIR) dynamics happens on an edge-colored graph with $N$ nodes~\cite{Pastor-Satorras2015, ramsey1987problem}. For simplicity, \change{in most of our analysis} we assume that there are only two layers of interactions. To have a concrete example in mind, think of a population of college students, where layer $\ell=1$ represents interactions in housing facilities, and layer $\ell=2$ represents classroom interactions. Individual layers of the network have pre-imposed block/community structure constructed according to the rules of the stochastic block model~\cite{holland1983stochastic}. Once more for simplicity, we assume the model to be homogeneous both in terms of degree and block sizes. Specifically,  nodes' degrees in layer $\ell$ obey a Poisson distribution with average  $\langle k^{(\ell)} \rangle$. Also, each individual $i$ is associated to a group $\sigma_i^{(\ell)} = 1, \ldots, Q^{(\ell)}$ in layer $\ell$, with all groups having the same size $q^{(\ell)}$. The mixing parameter $\mu$ quantifies the fraction of connections that each node has towards other nodes outside its own group. Please note that we use the same mixing parameter for both layers just for convenience, but this is not a requirement of our model. The correlation between the community structure of two layers is measured in terms of normalized mutual information (NMI)~\cite{danon2005comparing}. Perfectly aligned communities generate large NMI values, whereas uncorrelated block structures have low NMI values.
Concerning SIR dynamics, the recovery rate is set equal to one, whereas spreading events occur in both layers at rate $\beta$. The choice of using identical spreading rates for both layers is made for simplicity; the effective rate of spreading is anyway also a function of $\langle k^{(\ell)} \rangle$.
Initial conditions of the dynamics are such that all nodes are in the susceptible state, except for a single random node that is set in the infected state.

In the example of Fig.~\ref{fig:0}, we play only with one ingredient, that is the correlation of the community structure of the layers. Individual layers have a neat block structure, as they are generated for $\mu = 0.025$. However, the two structures are in one case maximally correlated (NMI = $1.0$) and in the other case completely uncorrelated (NMI = $0.0$).
In the example of the population of college students, correlated blocks occur when the conditional probability for two students to attend the same class given that they live in the same dorm is higher than the probability of two random students to attend the same class; uncorrelated community structure indicates that the two probabilities are equal.

The effect that correlation has on the outcome of SIR  spreading is particularly interesting even for a small population like the one considered in Fig.~\ref{fig:0}. Outbreaks are comparable in size between the correlated and uncorrelated cases; however, graphs with correlated community structure display milder and longer epidemics than graphs with uncorrelated groups.
\change{In essence, topological correlations among the layer-wise community structure of the network may have an impact on how the epidemic unfolds, but not on its magnitude. This is a fundamentally important aspect that was not captured in the analysis by Fan {\it et al.}~\cite{Fan2019}.}

\begin{figure*}[!htb]
    \centering
    \includegraphics[width=0.85\textwidth]{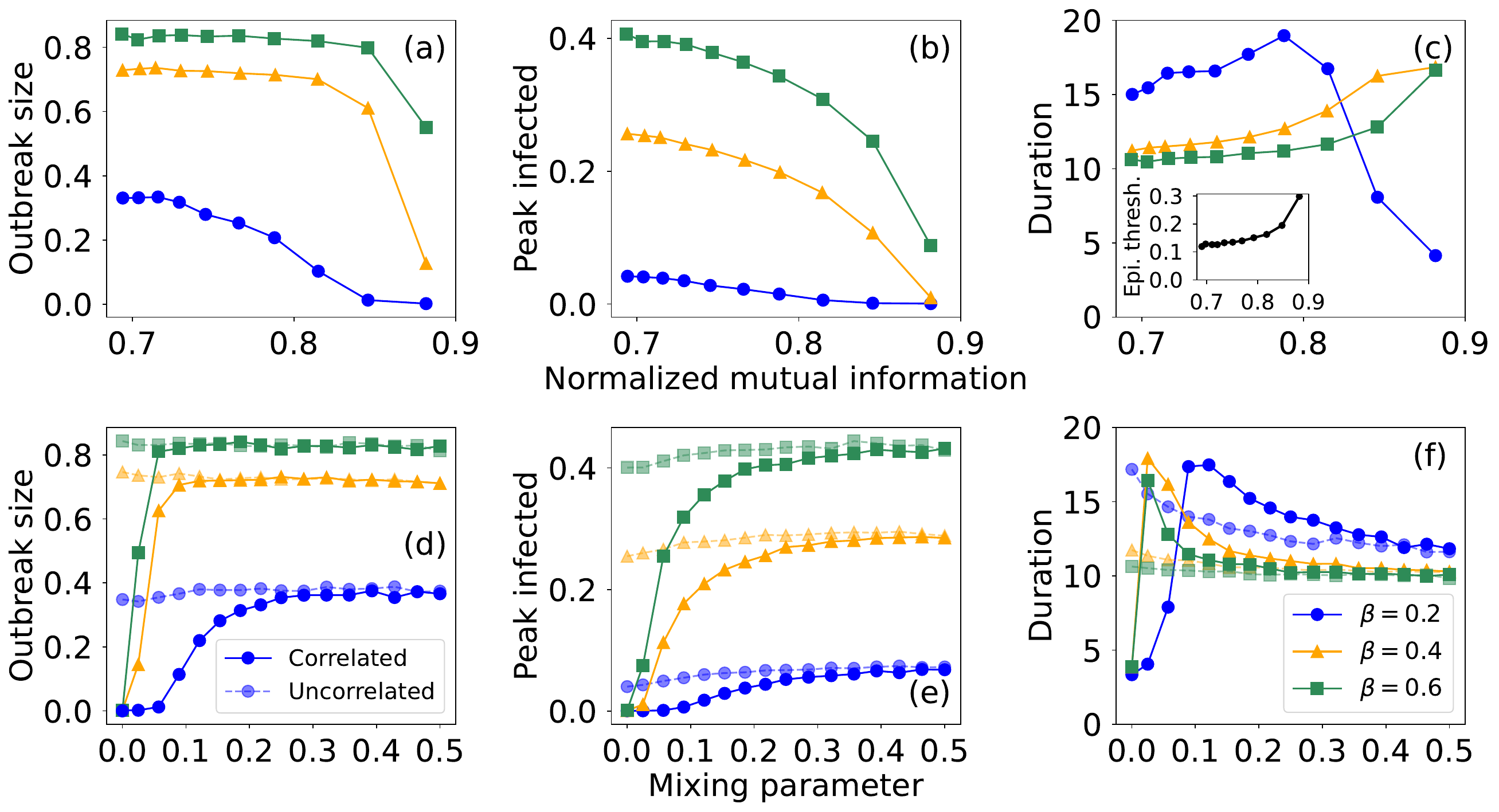}
    \caption{{\bf Epidemic spreading in synthetic group-structured populations.}
    (a) We consider edge-colored graphs with $N=10,000$ nodes and parameters $\langle k^{(1)} \rangle = 3$, $q^{(1)} = 5$,  $\langle k^{(2)} \rangle = 10$, $q^{(2)} = 25$, and $\mu = 0.025$. We tune the correlation among the community structure of the layers by swapping community memberships of nodes as explained in the Methods section. We simulate SIR dynamics, and measure the size of the outbreak. We plot it as a function of the NMI between the layers' partitions. Results are averaged over $V=5,000$ repetitions. Different colors/symbols refer to results valid for different choices of the spreading rate $\beta$, i.e., $\beta =0.2, 0.4$ and $0.6$, all corresponding to supercritical spreading rates [see inset in panel (c)]. \change{The values of the reproduction number are $R_0= 2.6$, $5.2$, and $7.8$, respectively.} (b) Same as in (a), but for the peak value of the fraction of infected. (c) Same as in (a), but for the duration of the epidemic.
    (d) We consider only the configurations corresponding to maximum (full curves, solid symbols) and minimum (dashed curves, transparent symbols) correlation among layers' partitions, and generate networks with variable mixing parameter $\mu$. We plot the size of the outbreak as function of $\mu$. Results represent averages over $V=5,000$ realizations of the model. (e) Same as in (d), but for the peak of the fraction of infected. (f) Same as in (d), but for the average duration of the epidemic.
    }
    \label{fig:3}
\end{figure*}


We study in a systematic manner the effect that the correlation among the community structure of the graph layers has on SIR spreading. To this end, we employ a simple algorithm that allows us to generate block structures anywhere between the maximally correlated and the uncorrelated configurations, see Methods for details. In Fig.~\ref{fig:3}, we display results for a graph with $N=10,000$ nodes. Model parameters are: $\langle k^{(1)} \rangle = 3$, $\langle k^{(2)} \rangle = 10$, $q^{(1)} = 5$, $q^{(2)} = 25$, and $\mu = 0.025$. We consider three distinct values of the SIR spreading rate $\beta = 0.2, 0.4$ and $0.6$, all of them corresponding to the endemic regime of the dynamics. As the figure clearly displays, increasing the correlation between the block structure of the layers reduces the severity of the epidemic. There is no apparent reduction in terms of outbreak size, unless the layer community partitions are very close to the configuration corresponding to maximum correlation. Severity reduction is instead mostly visible in terms of intensity and duration. Essentially, spreading slows down as the correlation of the partitions is increased, with a clear reduction in the peak value of the fraction of infected and an apparent increase in the total duration of the epidemic.

The above findings are valid for network layers that are sufficiently modular. When the strength of the community structure varies, nothing happens if the layers' partitions are uncorrelated, see Figs.~\ref{fig:3}d-f. However, if partitions are correlated, we observe an increase of epidemic severity as the community structure progressively becomes loose.
Further, we observe an interesting trend in the metrics of epidemic severity as the number/size of communities is varied (\change{Figs.~S1 and~S2}). The trend is visible only if partitions are correlated, and community structure is sufficiently strong. In such a case, the outbreak size, the duration and the peak of infected nodes grow as the size of the blocks decreases. Once more, if partitions are not correlated, then no change in the values of metrics of epidemic severity is visible as the size/number of clusters is varied.
\change{All the above considerations are still valid even if we consider models with heterogeneous community sizes (Fig.~S3) and node degrees (Fig.~S4). Further, results for edge-colored graphs composed of three layers are almost identical to those obtained on two-layer edge-colored graphs (Fig.~S5).}

\begin{figure*}[!htb]
    \centering
   \includegraphics[width=0.85\textwidth]{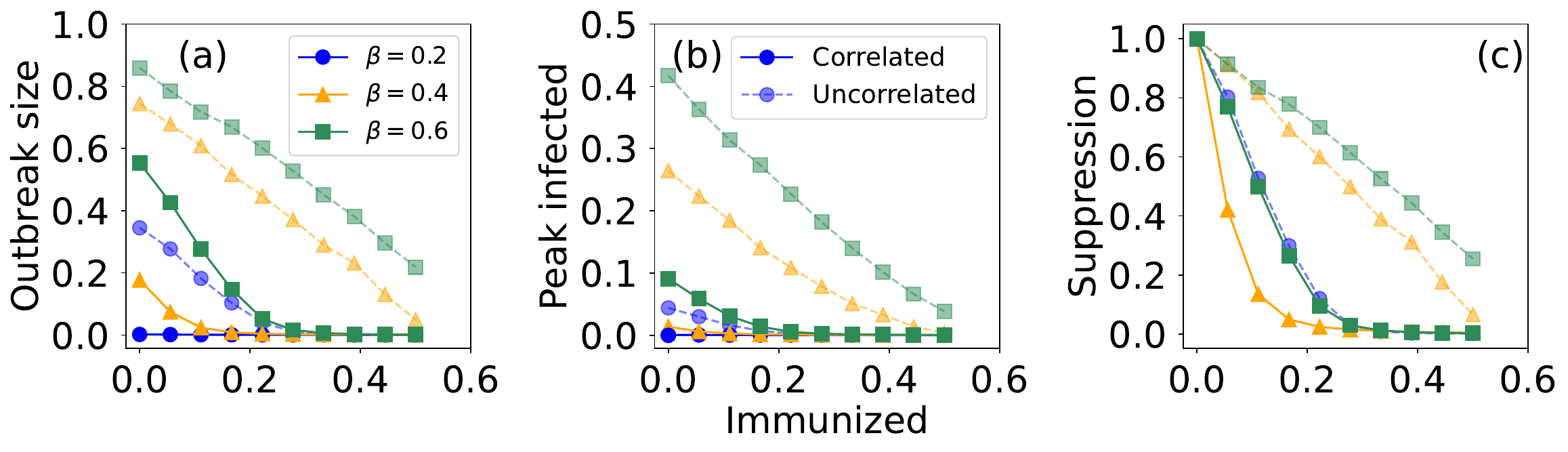}
   \caption{{\bf Immunization in synthetic group-structured populations.} (a) We consider the same experimental setting as in the bottom row of Fig.~\ref{fig:3}. \change{We set the mixing parameter $\mu=0.025$ and consider three values of the spreading rate $\beta$, i.e., $0.2$, $0.4$, and $0.6$, corresponding to the reproduction number $R_0 = 2.6$, $5.2$, and $7.8$, respectively}. We change, however, the initial condition of the dynamics, by immunizing a random fraction nodes. We then plot the size of the outbreak as a function of the fraction of immunized nodes. Results are obtained by averaging the outcome of $V=5,000$ repetitions of the epidemic process. (b) Same as in (a), but for the peak fraction of infected nodes. (c) We rescale the abscissa values of panel (a) by the outbreak size that is observed when a null fraction of nodes is immunized.}
    \label{fig:5}
\end{figure*}

Also, we analyze how the two main ingredients of our network model, i.e., the strength of the community structure of the individual layers and the correlation among the community partitions of the layers, influence the effectiveness of immunization in suppressing an epidemic, see Fig.~\ref{fig:5}. To this end, we simply change the initial conditions of the dynamics by imposing that a fraction of randomly chosen nodes is set to the recovered state. All other nodes are in the susceptible state, except for one randomly chosen initial spreader that is set in the infected state. The effectiveness of random immunization is greatly enhanced when communities are correlated: at parity of spreading rate $\beta$, correlating the group structures leads to 
even a five-fold 
reduction in outbreak size and peak of infection compared to the uncorrelated case; also, the overall behavior of the system obtained for $\beta=0.6$ in presence of group correlation is almost identical to the one observable for $\beta=0.2$ in a population with uncorrelated group structure.
A similar conclusion is reached by measuring the site-percolation thresholds of the networks, see 
\change{Fig.~S6.}
The site-percolation model is closely related to immunization,
as the immunized nodes and their connections are 
effectively
removed from the system and the spreading dynamics is governed by the network formed by the edges between the non-immunized nodes. 

All the above results are obtained by means of numerical simulations. In the Supplemental Material (SM), we describe the derivation of two theoretical approximations: the individual-based  mean-field approximation (IBMFA) and the group-based mean-field approximation (GBMFA). IBMFA is a standard theoretical approach that uses as input the topology of the network~\cite{Pastor-Satorras2015}. GBMFA is \change{an approximation introduced in this paper,} inspired by the so-called degree-based mean-field approximation where nodes with identical degrees are treated as indistinguishable elements~\cite{Pastor-Satorras2015}. The main difference in GBMFA is that classes of indistinguishable nodes are defined based on the layer-wise community structures.  IBMFA and GBMFA generate similar predictions; one of these predictions is that the epidemic threshold equals $\beta_c = \left( \langle k^{(1)} \rangle + \langle k^{(2)} \rangle \right)^{-1} $ for any level of correlation existing between the layers' community structures \newchange{, see Fig.~S7}. \change{Qualitatively, this theoretical prediction supports our numerical findings on the weak dependence of the long-term metrics of epidemic severity from topological correlations among the network layers, highlighting  once more fundamental differences existing between our results and those by Fan {\it et al.}~\cite{Fan2019}. From the quantitative point of view, however,} we find the mean-field approximations to be accurate enough only for populations with sufficiently large communities \change{or} sufficiently large average degree, see \newchange{Figs.~S8-S11}. These regimes, however, are of little importance in realistic settings where groups are generally much smaller than the size of the entire population and the average degree of individual nodes is small.


\begin{figure*}[!htb]
    \centering
    \includegraphics[width=0.85\textwidth]{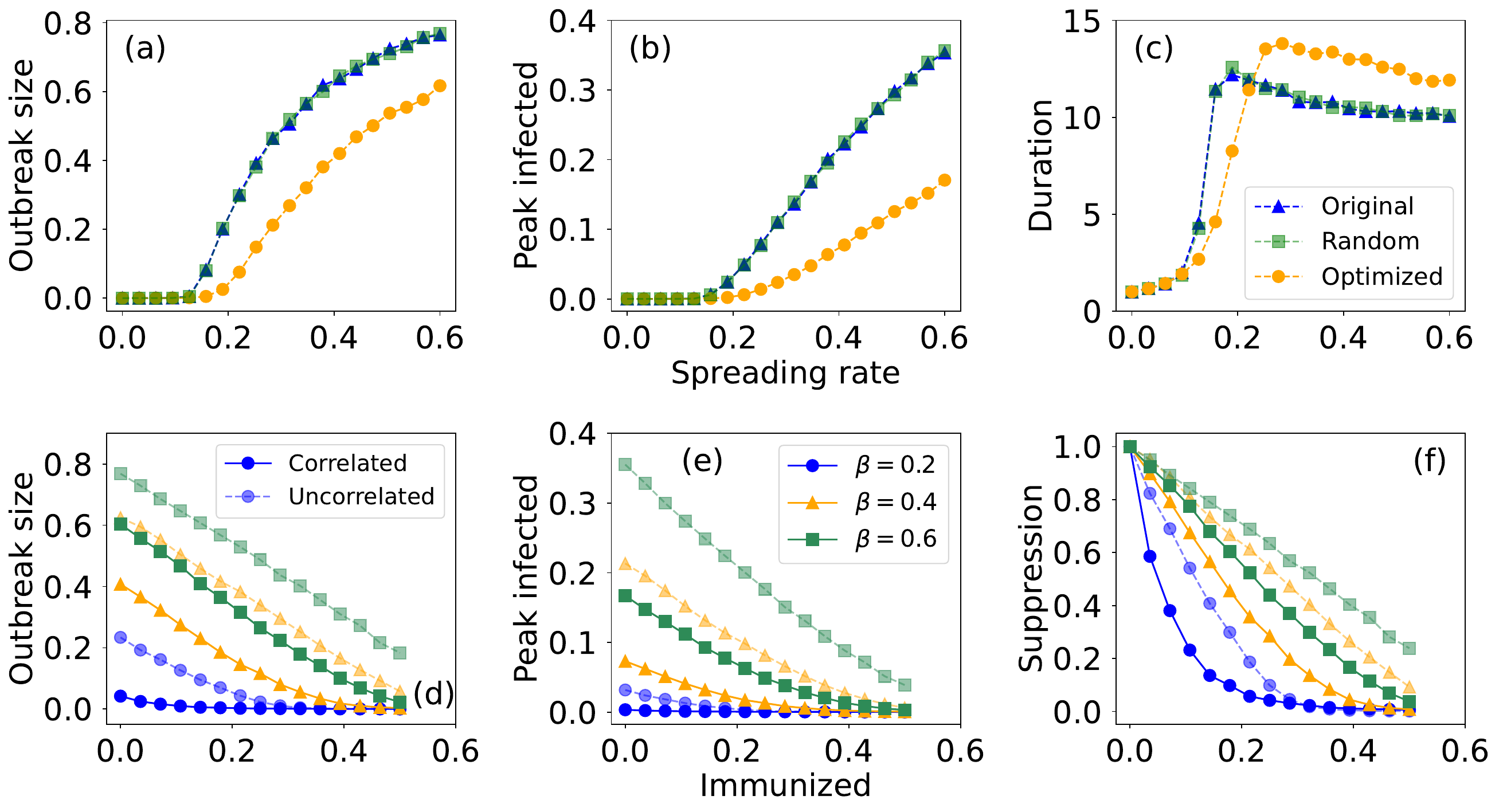}
    \caption{{\bf Epidemic spreading in the student population of the Indiana University Bloomington.}
    (a) We use data about housing and attendance for the Fall 2019 semester at the Indiana University Bloomington (IUB) campus to generate edge-colored graphs with block structure. The community partition in one layer reflects housing assignments; the partition in the other layer serves to group students based on their program and education level.  Different graphs are generated depending on whether network partitions are (i) those directly observed from the data, (ii) randomized, or (iii) optimized for maximum correlation. We then simulate SIR dynamics on the graphs and measure the average size of the outbreak as a function of the spreading rate $\beta$. Results are averaged over $V=5,000$ repetitions. (b) Same as in (a), but for the peak fraction of infected. (c) Same as in (a), but for the duration of the spreading process. (d) We plot the size of the outbreak as a function of the fraction of individuals that are initially immunized. \change{We consider three values of the spreading rate $\beta$, i.e., $0.2$, $0.4$, and $0.6$, corresponding to the reproduction number $R_0= 2$, $4$, and $6$, respectively.} Different symbols correspond to different $\beta$ values; full curves and solid symbols indicate the optimized configuration considered in panel (a); dashed curves and transparent symbols refer to graphs created using ground-truth partitions.  Results are averaged over $V=5,000$ repetitions. (e) Same as in (d), but for the peak fraction of infected. (f) Same as in (d), but with abscissa values rescaled by the outbreak size observed when zero individuals are initially immunized.
}
    \label{fig:real}
\end{figure*}

\subsection{College housing/attendance network}

The paper has thus far presented simulations on synthetic 
group-structured populations aimed at understanding the fundamental principles of epidemic spreading 
in these systems. In this section, we demonstrate the significance and effectiveness of 
these principles in a real-world scenario. 
We examine data about housing and attendance
at Indiana University Bloomington (IUB), see Methods for details. 
There are $N=10,132$ individuals in the IUB dataset, each representing a 
student who resided in one of the campus facilities during the Fall 2019 semester. We use layer $\ell=1$ to represent contacts between students in housing settings. We form $Q^{(1)} = 396$ groups composed of students living in the same floor of a large dormitory, or in the same Greek house. The average size of these groups is $\langle q^{(1)} \rangle = 25.58$. Layer $\ell =2$ is used to represent classroom interactions. We form $Q^{(2)} = 600$ groups of students based on their enrollment program (e.g., Computer Science, Finance, Mathematics, Physics) and their education level (from Freshman to Ph.D.). The average size of these groups is $\langle q^{(2)} \rangle = 16.86$. \change{Note that the partitions within the classroom and residential layers are disjoint. Moreover, 
the group sizes in the two partitions are not homogeneous, \newchange{see Fig.~S12.}}

The ground-truth community partitions display a correlation level similar to that of the uncorrelated case. Specifically, the NMI between the partitions obtained from the data is $0.389$, which is comparable with the one obtained by randomizing the dormitory assignment of students which is $0.327$. We also create an artificial configuration, where students are re-assigned to residences so that the NMI between the partitions of the two layers is maximized. The optimization technique we employ is described in the Methods section. The optimized configuration corresponds to the NMI value $0.934$.

Given the three configurations described above, we generate modular network layers using the Lancichinetti-Fortunato-Radicchi (LFR) model~\cite{lancichinetti2008benchmark}, see Methods for details. Nodes' degrees obey Poisson distributions with average equal to $\langle k^{(1)} \rangle = \langle k^{(2)} \rangle = 5$. The strength of the community structure is controlled by the value of the mixing parameter, set here to $\mu = 0.025$. We stress that the sizes of the communities in the two layers is always set equal to the one learned directly from the real data.

In Figs.~\ref{fig:real}a-c, we display the metrics of epidemic severity as functions of the spreading rate. No difference between the ground-truth and randomized configurations is noticeable. The optimized configuration displays instead a less severe epidemic compared to the other two configurations.
Also, we notice that random immunization becomes more effective when communities are correlated than when no correlation among the layers' partitions is present, Figs.~\ref{fig:real}d-f. 
\change{Qualitatively similar results are obtained also for $\mu = 0.1$, see \newchange{Fig.~S13.}}

\section{Discussion}

In this paper, we studied susceptible-infected-recovered (SIR) spreading in group-structured populations. The real system that inspired our work is a population of college students that reside on campus during an academic semester. Recurrent interactions among students are due to group activities/settings, e.g., housing and class enrollment; each setting is responsible for the formation of network layers 
with modular structure. At the level of individual layers, modular structure is characterized by the strength and size of the blocks/communities that are present in the layer. In the example of the population of college students, these correspond to the size of classes or dorms, and to the level of mixing between students depending on their enrollment or housing assignment. Correlation among the layer-wise community structures is indicative instead for the likelihood of two nodes to belong to the same community in both layers, i.e., the conditional probability of two students to attend the same class given that they live in the same dorm. All these parameters play fundamental roles for the determination of properties characterizing the spread of an infectious disease within the population.  

In the first part of the paper, we considered synthetic populations where the above ingredients can be finely tuned. By means of extensive numerical simulations, we showed that SIR dynamics is characterized by mild and long spreading processes if the communities of the individual layers are neat, and the correlation among the layer-wise community structure is sufficiently high. On contrary, weak and/or uncorrelated community structures generally correspond to short and intense outbreaks. Immunization in populations with correlated communities is also greatly more effective than in populations with uncorrelated layer-wise community structures. 

In the second part of the paper, we took advantage of real data about housing/attendance of students at the Indiana University Bloomington campus, and numerically simulated SIR dynamics in this population. Ground-truth data correspond to layer-wise partitions that are uncorrelated. By appropriately re-assigning students to housing facilities on the basis of their enrollment profiles, we were able to show that the severity of an epidemic can be significantly reduced compared to the case in which no re-assignment is performed. The re-assignment strategy also enhances the effectiveness of random immunization in suppressing disease spreading.

\change{The above results are obtained under the assumption that the layer-wise group structure is sufficiently neat, i.e., for sufficiently low values of the mixing parameter $\mu$. This assumption can be empirically justified by analyzing publicly available datasets concerning contact networks within group-structured populations. First, we consider a contact network for a primary school. The dataset consists of events of physical proximity between students and teachers~\cite{stehle2011high}. Each event of physical proximity is associated to a duration. The system contains $242$ individuals and $11$ communities ($10$ classrooms + teachers). Two days of observation are included in the dataset: day one consists of $37,414$ contact events and day two of $40,188$ contact events. We find that the value of the weighted mixing parameter $\langle \mu \rangle $, i.e., the average fraction of contact time spent by the students outside their classrooms, are $\langle \mu \rangle = 2.3 \times 10^{-5}$ and $\langle \mu \rangle = 2.6 \times 10^{-5}$, respectively. Next, we look at a dataset concerning physical proximity in an office building~\cite{genois2015data}. The dataset consists of $9,827$ contact events between $92$ individuals organized in $5$ departments (communities). We find $\langle \mu \rangle = 9.1 \times 10^{-4}$.
}

Our findings stem from the basic principle that the severity of an epidemic is proportional to the overall level of mixing between individuals of a population. Correlated modular structure in our framework corresponds to redundant interactions among subsets of individuals, which speeds up the propagation of a disease within a subset, but also slows down the diffusion across subsets thus in the overall population.  Although having a radically different implementation, the mechanism of reducing the severity of an epidemic by enhancing group correlation is similar to the one at the basis of the mitigation strategy proposed by Meidan {\it et al.}, consisting of an alternating lockdown policy that involves half of the population at a time~\cite{meidan2020alternating}. 

A very important remark is that our findings are apparent only from numerical simulations of SIR dynamics, but they do not emerge from the numerical integration of mean-field equations aimed at approximating SIR dynamics. The mismatch between ground-truth dynamics and its approximations is particularly apparent in realistic settings where groups are small compared to the population size; also, the mismatch mostly concerns early- and mid-stages of SIR dynamics, rather than its long-term features.

All our results are based on important simplifications concerning the spreading dynamics of realistic diseases as well as the contact networks that are the basis of the spread of infectious diseases within real-world populations. Nonetheless, we believe they provide some easy-to-implement principles to reduce the severity of real epidemics that managers can take under consideration when planning group activities in schools, colleges or other large organizations.

\section{Methods}

\subsection{Edge-colored block-structured graphs}

We consider edge-colored graphs where $N$ nodes are connected via \change{$L$} different types or layers of interactions~\cite{ramsey1987problem}. \change{We use $L=2$ in most of our analysis; some results for $L=3$ are presented in the SM.}
The topology of the layer 
\change{$\ell = 1, 2, \ldots, L$}
is fully specified by the adjacency matrix $A^{(\ell)}$. The generic element of this matrix $A^{(\ell)}_{ij} = A^{(\ell)}_{ji} =1$ if nodes $i$ and $j$ are connected, and $A^{(\ell)}_{ij} = A^{(\ell)}_{ji} =0$
otherwise. 

\subsubsection*{Synthetic graphs}

In \change{the majority of} our experiments on synthetic graphs, network layers are instances of the version of the stochastic block model (SBM)~\cite{holland1983stochastic} named planted partition model~\cite{condon1999algorithms}. 
Inputs for the generation of the network layer $\ell$ are the average degree $\langle k^{(\ell)} \rangle$, the mixing parameter $\mu^{(\ell)}$, the size $q^{(\ell)}$ of each community, and the vector $\vec{\sigma}^{(\ell)} = (\sigma_1^{(\ell)}, \sigma_2^{(\ell)}, \ldots, \sigma_i^{(\ell)}, \ldots,   \sigma_N^{(\ell)})$ denoting the community membership of individual nodes, i.e., node $i$ belongs to community $\sigma_i^{(\ell)} = 1, \ldots, Q^{(\ell)}$, with $Q^{(\ell)}$ total number of blocks in layer $\ell$; the output is the adjacency matrix $A^{(\ell)}$. The element $A^{(\ell)}_{ij}$ is a Bernoulli random variate equal to one with probability $p^{(\ell)}_{in} = \langle k^{(\ell)} \rangle (1-\mu^{(\ell)}) / (q^{(\ell)}-1)$ if $\sigma_i^{(\ell)} = \sigma_j^{(\ell)}$, and with probability 
$p^{(\ell)}_{out} = \langle k^{(\ell)} \rangle \mu^{(\ell)} /(N-q^{(\ell)})$ if $\sigma_i^{(\ell)} \neq \sigma_j^{(\ell)}$. \change{Please note that this is the only model used to generate edge-colored graphs with $L=3$ layers.}

\change{In some of our experiments on synthetic graphs, we generate network layers using the Lancichinetti-Fortunato-Radicchi (LFR) model~\cite{lancichinetti2008benchmark}, i.e., a variant of the SBM designed to deal with heterogeneous community sizes and/or nodes' degrees. 
More specifically, in the set of experiments where we consider power-law distributed community sizes, we first generate the community sizes $q^{(\ell)}_1, q^{(\ell)}_2, \ldots, q^{(\ell)}_{Q^{(\ell)}}$. These are integer numbers randomly extracted from the distribution $P(q) \sim q^{-\tau}$ for $q \in [q_{\min}, q_{\max}]$ and $P(q) = 0$ otherwise, with $\tau$, $q_{\min}$ and $q_{\max}$ input parameters of the model. Clearly, we have that $\sum_{i=1}^{Q^{(\ell)}} q_i^{(\ell)} = N$. Please note that the same sequence for the community sizes is used for all layers of the network. Nodes are assigned to communities according to their sizes to form the community vector $\vec{\sigma}^{(\ell)}$. For the generation of the topology of the graph layer $\ell$, we first set the degree of each node to be a random variate extracted from a Poisson distribution with average $\langle k^{(\ell)} \rangle$. We then connect pairs of nodes at random, taking care of preserving for each node its pre-assigned degree value, as well as the ratio $\mu^{(\ell)}$ between inter-community and total connections. Here the average degree $\langle k^{(\ell)} \rangle$ and the mixing parameter $\mu^{(\ell)}$ are input parameters of the model. In the set of experiments where we consider power-law degree distributions, the size of all communities in layer $\ell$ is equal to $q^{(\ell)}$, an input parameter of the model. We then assign degrees to nodes by extracting random numbers from the power-law distribution $P(k) \sim k^{-\gamma}$ for $k \in [k_{\min}, k_{\max}]$ and $P(k) = 0$ otherwise, with $\gamma$, $k_{\min}$ and $k_{\max}$ input parameters of the model. The degree sequence of each layer is generated independently. We then connect pairs of nodes at random, taking care of preserving for each node its pre-assigned degree value, as well as the ratio $\mu^{(\ell)}$ between inter-community and total connections, with 
 the mixing parameter $\mu^{(\ell)}$ being an input parameter of the model.
}

\subsubsection*{College housing/attendance network}

Also, we consider a dataset containing information about class attendance of students who resided in the Indiana University Bloomington (IUB) campus during the Fall 2019 semester~\cite{deom2021data}. The dataset comprises $N = 10,132$ students. We use layer $\ell=1$ to represent housing, and group together students living in the same floor of a dormitory building or in the same Greek house. We form $Q^{(1)} = 396$ groups of average size $\langle q^{(1)} \rangle = 25.58$. Please note that the size of the various groups is not constant, see \newchange{Fig.~S12}. We use layer $\ell=2$ to represent class interactions. Each student in the dataset is associated to one of the $218$ degree programs (e.g., Law, Mathematics, Physics) and $6$ education levels (i.e., Freshman, Sophomore, Junior, Senior, Master, and Ph.D.). On the basis of our data, we form $Q^{(2)}  = 600$ blocks of average size $\langle q^{(2)} \rangle = 16.86$, by grouping together students belonging to the same degree program and education level. Note that not all possible compartments defined by the pairs degree program/education level contain at least one student. Also for layer $\ell=2$, the size of the various groups is not a constant, see \newchange{Fig.~S12}.  
Network layers considered in the analysis of the IUB dataset are generated according to the
\change{LFR model.}
Inputs required for the generation of the topology of layer $\ell$ are the average degree $\langle k^{(\ell)} \rangle$, the mixing parameter $\mu^{(\ell)}$, and the community-membership vector $\vec{\sigma}^{(\ell)}$. For the generation of the topology of the graph layer $\ell$, we first set the degree of each node to be a random variate extracted from a Poisson distribution with average $\langle k^{(\ell)} \rangle$. We then connect pairs of nodes at random, taking care of preserving for each node its pre-assigned degree value, as well as the ratio $\mu^{(\ell)}$ between inter-community and total connections.

\subsection{Measuring and tuning correlation among the community structures of the layers}

We measure the similarity between the community structure of two network layers, namely $\vec{\sigma}^{(1)}$ and $\vec{\sigma}^{(2)}$, in terms of normalized mutual information (NMI)~\cite{danon2005comparing}. By definition, NMI values are in the range $[0, 1]$; however, the actual minimum and maximum values that NMI can assume
depend on the size and number of clusters in the two partitions. Small values of the NMI are associated to uncorrelated partitions, whereas large NMI values indicate correlated partitions.

\subsubsection*{Synthetic graphs}

In our systematic tests on synthetic network models \change{that are constructed using the SBM}, we generate perfectly correlated partitions as follows. First, we fix the size $q^{(1)}$ of the clusters in layer $\ell = 1$; then we determine the size of the clusters in layer $\ell = 2$ as $q^{(2)} = d\, q^{(1)}$, where $d$ is a tunable integer parameter. For simplicity, we appropriately choose $q^{(1)}$ and $d$ so that both $q^{(1)}$ and $q^{(2)}$ are divisors of $N$. For example, in many of our tests we have $N=10,000$, $q^{(1)} =5$, $d=5$, and $q^{(2)} = 25$. Finally, we set $\sigma_i^{(\ell)} = \lfloor i /  q^{(\ell)} \rfloor + 1$ for all $i=1, \ldots, N$ and for $\ell = 1, 2$, with $\lfloor \cdot \rfloor$ floor function.

\change{In the tests where we take advantage of the LFR model, the size of the communities is identical in both layers. We thus generate perfectly correlated partitions by setting $\sigma_i^{(1)} = \sigma_i^{(2)}$ for all nodes $i$ in the network.}

Starting from the partitions \change{$\vec{\sigma}^{(1)}$ and $\vec{\sigma}^{(2)}$} that are maximally correlated, their NMI is reduced by considering each node, and swapping with probability $r$ its community membership in layer $\ell=1$ with that of another randomly selected node. For $r=0$, no swaps are actually performed so that the correlation of the community partitions is preserved; for $r=1$, community labels in layer $\ell = 2$ are completely randomized compared to the initial configuration so that correlation between the layers' community memberships becomes minimal.

\change{In the set of experiments where we consider edge-colored graphs composed of $L=3$ SMB-generated network layers, all communities have the same size thus we set $\sigma_i^{(1)} = \sigma_i^{(2)} = \sigma_i^{(3)}$ for all nodes $i$ in the network to generate perfectly correlated partitions. 
To decrease correlation, we consider each node and swap with probability $r$ its community membership in layers $\ell=1$ and $\ell=2$ with that of another randomly selected node. Please note that we only consider the cases $r=0$ and $r=1$ for this specific sets of experiments.}

\subsubsection*{College housing/attendance network}

To create completely uncorrelated partitions, we simply swap the housing assignment of each student with that of another randomly selected student.

To create a configuration with highly correlated partitions, we take advantage of the following greedy optimization technique. We define the tuple $\mathcal{P}^{(\ell)} = \left( z_1^{(\ell)} , \ldots, z_{Q^{(\ell)}}^{(\ell)} \right)$, 
\change{where $z_v^{(\ell)}$ denotes the remaining availability of group $v$ in layer $\ell$. }
We initially set $z_v^{(\ell)} = q_v^{(\ell)}$ for $v = 1, \ldots, Q^{(\ell)}$, where $q_v^{(\ell)}$\change{, i.e., the size of group $v$ in layer $\ell$,} is imposed by the IUB housing/attendance data. We set $M=0$ denoting the number of elements that have been assigned to groups in the layers. We then iterate the following operations:

\begin{enumerate}

\item We find the group label $r^{(\ell)}$ corresponding to the largest element in $\mathcal{P}{(\ell)}$.
\change{We can either have $r^{(1)} \geq r^{(2)}$, or $r^{(2)} \geq r^{(1)}$. We indicate with $r^{(\ell_t)}$ the larger of the two values, where $\ell_t$ is the label of the corresponding layer; we use $r^{(\ell_s)}$ to indicate the smaller of the two values, with $\ell_s$ denoting the label of the corresponding layer.} 

\item We set $\sigma_i^{(\ell_s)} = r^{(\ell_s)}$ and $\sigma_i^{(\ell_t)} = r^{(\ell_t)}$ for all $i=M+1, \ldots, M+z_{r^{(\ell_s)}}^{(\ell_s)}$,
\change{i.e., we assign $z_{r^{(\ell_s)}}^{(\ell_s)}$ nodes to coherent groups in both layers.}

\item We update $M \to M + z_{r^{(\ell_s)}}^{(\ell_s)} $, $z_{r^{(\ell_t)}}^{(\ell_t)} \to z_{r^{(\ell_t)}}^{(\ell_t)} - z_{r^{(\ell_s)}}^{(\ell_s)}$ and $z_{r^{(\ell_s)}}^{(\ell_s)} \to 0$. 

\item \change{Until all the individuals have been assigned to a group, i.e., }$M<N$, we go back to point 1, otherwise, we end the iterative procedure.

\end{enumerate}

In the above step 2, we maximize the overlap of the largest groups still available in the layers. 
This greedy choice is iterated multiple times until all individuals have been assigned to groups in both layers. At the end of the algorithm, all groups of a given layer will have the desired size.
Since the labels of the nodes are completely arbitrary, the obtained partitions can be thought as equivalent to re-assigning students to dorms based on their enrollment.

\subsection{SIR dynamics on edge-colored graphs}

We consider continuous-time Susceptible-Infected-Recovered (SIR) dynamics on an edge-colored graph~\cite{Fan2019}. Without loss of generality, we assume that the rate of recovery equals one.
\change{This is a standard setting as the behaviour of SIR models is determined by the ratio of the spreading and recovery rates rather than their raw values.} We indicate with $\rho_i(t)= S, I$ or $R$ the state of node $i$ at time $t$. Our initial conditions are generally such that all nodes are in the $S$ state, except for a randomly selected node $v$ that is in state $I$, i.e., $\rho_i(t=0) = S$ for $i = 1, \ldots, N$ and $i \neq v$, while $\rho_v(t=0) = I$. In some of our experiments, we also impose that a fraction $f$ of randomly selected nodes is in the recovered state, i.e., $\rho_i(t=0) = R$ with probability $f$ for all $i = 1, \ldots, N$.

Starting from an initial condition, the rules of SIR dynamics are as follows. A generic node $i$ can change its state in the infinitesimal time interval $[t, t+dt]$ if: (i) Node $i$ is in the infected state and recovers with probability $dt$, i.e., $\rho_i(t) = I \to \rho_i(t+dt) = R$ with probability $dt$; 
\change{(ii) Node $i$ is in the susceptible state and gets infected with probability $\beta^{(\ell)} dt$ by its infected neighbor $j$ on layer $\ell$, i.e., $\rho_i(t) = S \to \rho_i(t+dt) = I$ with probability $\beta^{(\ell)} dt$ if $\rho_j(t) = I$ and $A^{(\ell)}_{ij} =1$. Please note that spreading events of type of (ii) can happen for all individual edges in all layers in the graph, thus if node $i$ is in the $S$ state at time $t$, the overall probability to get infected in the time interval $[t, t + dt]$ is $\textrm{Prob.}(\rho_i(t) =S \to \rho_i(t+dt) = I) = 1 - \prod_{\ell=1}^L (1 - \beta^{(\ell)} \, dt)^{g_i^{(\ell)}}$, where  $g_i^{(\ell)} = \sum_j A_{ij}^{(\ell)} \delta (\rho_j(t), I)$ is the number of infected neighbors that node $i$ has in layer $\ell$ at time $t$ with $\delta (x, y) = 1$ if $x=y$ and $\delta (x, y) = 0$ otherwise.}
The dynamics proceeds until no nodes are in the infected state. The model is efficiently implemented via the Gillespie's algorithm~\cite{gillespie1977exact}.

\subsubsection*{Metrics of epidemic severity}

We characterize the behavior of the SIR model on a given edge-colored graph by measuring the values of the following metrics:

\begin{itemize}

\item[(i)] The outbreak size, i.e., the fraction of nodes that are in state $R$ at the end of the epidemic. 

\item[(ii)] The peak of infection, i.e., the maximum value of the fraction of infected nodes that are simultaneously present in the network.

\item[(iii)] The epidemic duration, i.e., the instant of time when the last infected node recovers.

\end{itemize}

Given an edge-colored graph and the parameters \change{$\beta^{(\ell)}$}
 of the epidemic model, we simulate multiple times SIR dynamics. Values of the above metrics displayed in our figures correspond to their sample averages.

 \change{In the inset of Fig.~\ref{fig:3}(c),}
we approximate the pseudo-critical epidemic threshold
 of the given edge-colored graph as the value of the spreading rate $\beta$ corresponding to the maximum of the ratio between the sample standard deviation and the sample average of the outbreak size.

 \change{
 \subsection{Code and data availability}
 Code and data used in this study are available at \url{https://github.com/SiddharthP96/Multiplex-Epidemic-Spreading}.
 }

\begin{acknowledgements}
This project was partially supported by the Army Research Office under contract number W911NF-21-1-0194, by the Air Force Office of Scientific Research under award numbers FA9550-19-1-0391 and FA9550-21-1-0446, and by the National Science Foundation under award numbers 1927418 and 1735095. The
funders had no role in study design, data collection and
analysis, the decision to publish, or any opinions, findings,
and conclusions or recommendations expressed in the
manuscript.
\end{acknowledgements}


%

\newpage
\clearpage

\renewcommand{\theequation}{S\arabic{equation}}
\setcounter{equation}{0}
\renewcommand{\thefigure}{S\arabic{figure}}
\setcounter{figure}{0}
\renewcommand{\thetable}{S\arabic{table}}
\setcounter{table}{0}
\setcounter{section}{0}
\renewcommand{\thesection}{S\arabic{section}}

\section*{Supplemental Material}

\subsection*{Community sizes and epidemic dynamics on edge-colored graphs}

\begin{figure*}[t]
   \centering
   \includegraphics[width=0.85\textwidth]{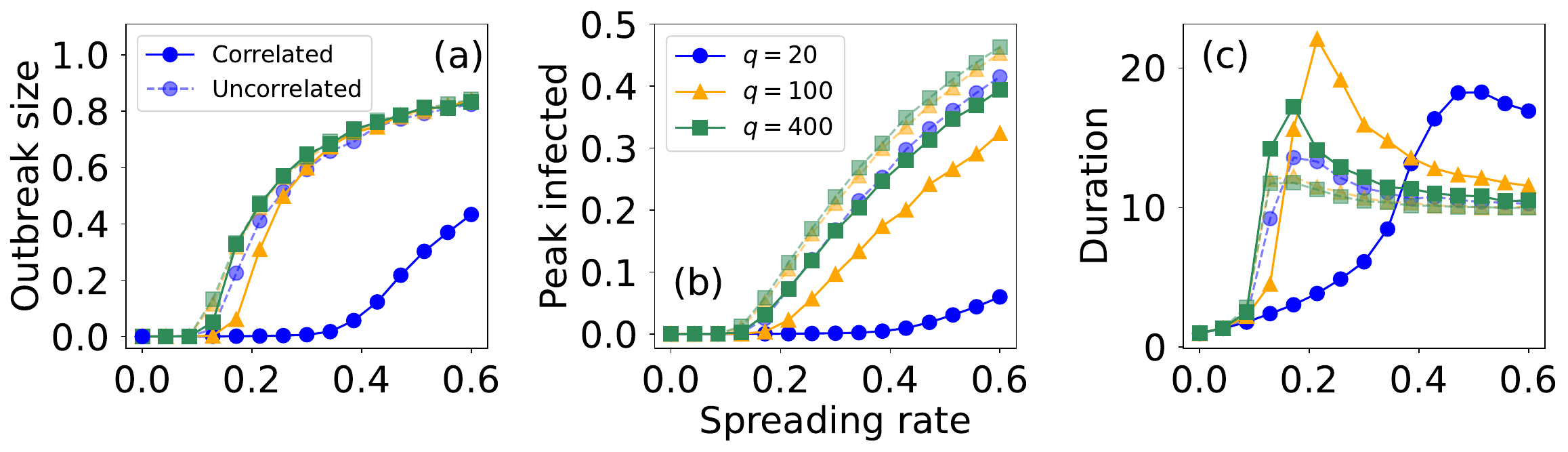}
   \caption{\textbf{Epidemic spreading in synthetic group-structured populations with variable group sizes.} We consider edge-colored graphs with $N=10,000$ nodes and label shuffling probabilities $r=0$ (correlated community structure) and $r=1$ (uncorrelated community structure). Graphs with $\langle k^{(1)} \rangle = \langle k^{(2)} \rangle = 6$ and $\mu = 0.025$ are considered. We set the community sizes $q^{(1)} = q^{(2)} = q$, and consider $q = 20, 100,$ and $400$. The solid and dotted curves show the indicated epidemic properties as a function of the spreading rate $\beta$ for edge-colored graphs with correlated and uncorrelated community structures, respectively. We run $V=5,000$ simulations of the SIR dynamics, and display average values in the plot. (a) Outbreak size, (b) fraction of the infected population at the peak, and (c) total duration of the outbreak are plotted as functions the spreading rate $\beta$.}
   \label{fig:groupsize}
\end{figure*}

\begin{figure*}[t]
   \centering
   \includegraphics[width=0.85\textwidth]{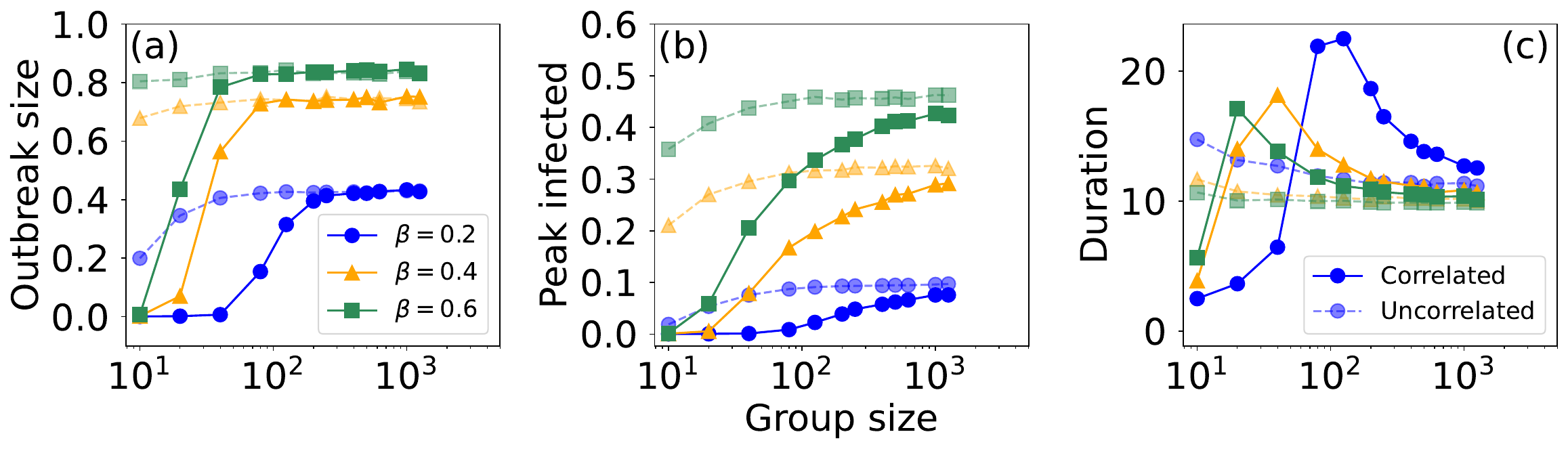}
   \caption{\change{\textbf{Epidemic spreading in synthetic group-structured populations with variable group sizes.} We consider edge-colored graphs with $N=10,000$ nodes and label shuffling probabilities $r=0$ (correlated community structure) and $r=1$ (uncorrelated community structure). Graphs with $\langle k^{(1)} \rangle = \langle k^{(2)} \rangle = 6$ and $\mu = 0.025$ are considered. We set the community sizes $q^{(1)} = q^{(2)} = q$, and vary $q$ from $10$ to $1250$. The solid and dotted curves show the indicated epidemic properties as a function of the group size $q$ for edge-colored graphs with correlated and uncorrelated community structures, respectively. We run $V=5,000$ simulations of the SIR dynamics, and display average values in the plot. (a) Outbreak size, (b) fraction of the infected population at the peak, and (c) total duration of the outbreak are plotted as functions of the spreading rate $\beta$.}}
   \label{fig:groupsize2}
\end{figure*}

In Fig.~\ref{fig:groupsize}, we vary the group size in synthetic edge-colored graphs and show its effect on the metrics of epidemic severity. We look at the outbreak size, the fraction of the population infected during the peak, and the duration in the correlated and uncorrelated edge-colored graphs as a function of the spreading rate $\beta$. We set the community sizes $q^{(1)} = q^{(2)} = q$ and consider three different values of $q$. We consider graphs with size $N=10,000$, mixing parameter $\mu=0.025$, and average degree $\langle k \rangle = 6$. \change{Furthermore, we consider the same setting and plot the outbreak size, the fraction of the population infected during the peak, and the duration in the correlated and uncorrelated edge-colored graphs as  functions of the group size $q$ in Fig.~\ref{fig:groupsize2}.}

\begin{figure*}[!htb]
   \centering
   \includegraphics[width=0.85\textwidth]{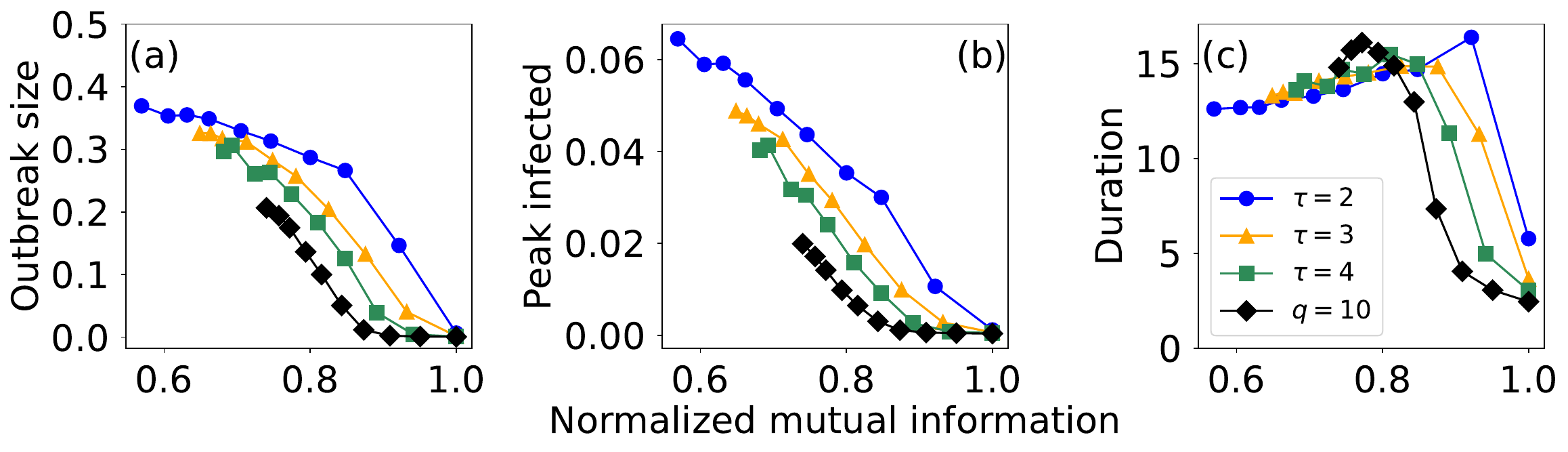}
     \caption{\change{\textbf{Epidemic spreading in synthetic group-structured populations with heterogeneous group sizes.} We consider edge-colored graphs with $N = 10,000$ nodes. Graphs with $\langle k^{(1)} \rangle = \langle k^{(2)}\rangle = 6$ and $\mu = 0.025$ are considered. We choose the community sizes $q$ from the power-law distribution $P(q) \sim q^{-\tau}$ for $q \in [10, 100]$ and $P(q) = 0 $ otherwise. We consider exponent values $\tau = 2$, $3$, and $4$. As a term of comparison, we also display results valid when communities have all identical size 
   $q=10$. We fix the spreading rate to $\beta=0.2$ corresponding to reproduction number $R_0=2.4$. We tune the correlation among the community structure of the layers by swapping community memberships of nodes as explained in the Methods section. We run $V = 5,000$ simulations of the SIR dynamics, and display average values in the plot. (a) Outbreak size, (b) fraction of the infected population at the peak, and (c) total duration of the outbreak are plotted as functions of the normalized mutual information between the two group structures.}}
   \label{fig:heterogroup}
\end{figure*}

\change{We also explore the effect of the heterogeneous community sizes on the outbreak. We consider edge-colored graphs with $N = 10,000$ nodes, average degrees $\langle k^{(1)} \rangle = \langle k^{(2)}\rangle = 6$ and $\mu = 0.025$.  We choose the community sizes $q$ from the power-law distribution $P(q) \sim q^{-\tau}$ for $q \in [10, 100]$ and $P(q) = 0 $ otherwise. We consider exponent values $\tau = 2$, $3$, and $4$. 
Community sizes $q_1, q_2, \ldots, q_Q$ are generated from the above power-law distribution and used in both layers to define their block structure.
As a term of comparison, we also display results valid when communities have all identical size 
   $q=10$. We tune the correlation among the community structure of the layers by swapping community memberships of nodes as explained in the Methods section. We run $V = 5,000$ simulations of the SIR dynamics, and display average values in Figure \ref{fig:heterogroup}}.

\change{\subsection*{Edge-colored graphs with heterogeneous degree distributions}}

\begin{figure*}[!htb]
   \centering
   \includegraphics[width=0.85\textwidth]{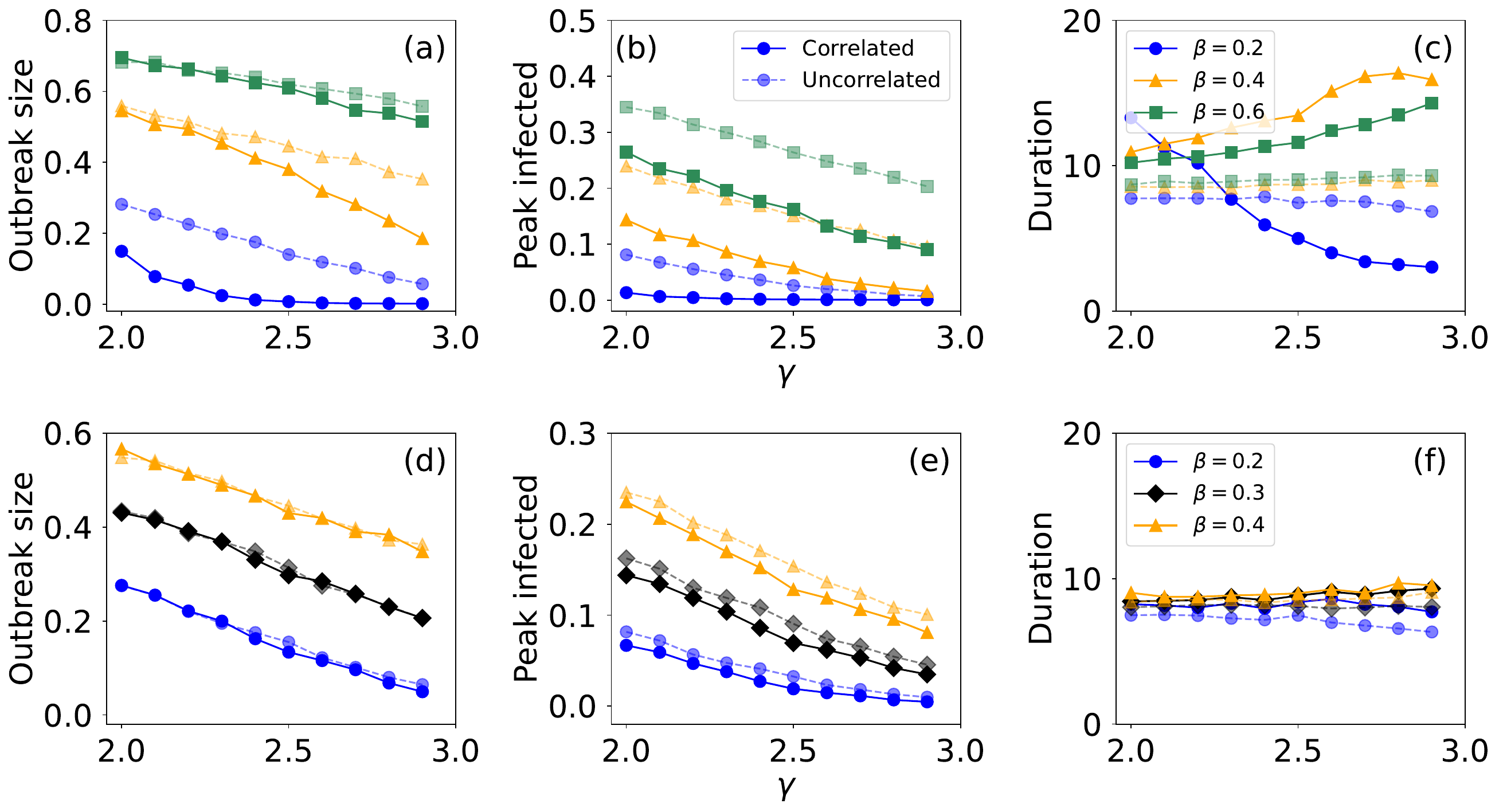}
   \caption{\change{\textbf{Epidemic spreading in synthetic group-structured populations with heterogeneous degree distributions.} (a) We consider edge-colored graphs with $N = 10,000$ nodes.  Nodes in the layers are organized in communities of size $q = 100$, and their degrees are chosen at random from a power-law degree distribution given by $P(k) \sim k^{-\gamma}$ for $k \in [2, \sqrt{q}]$ and $P(k) = 0 $ otherwise. The mixing parameter is $\mu = 0.025$. We consider the cases when the layer-wise community structures are maximally correlated (solid curves and solid symbols) and when they are uncorrelated (dotted curves and transparent symbols). We plot (a) the outbreak size, (b) the fraction of the infected population at the peak, and (c) the total duration of the outbreak as functions of the exponent $\gamma$.
   Different combinations of colors and symbols are obtained for different values of the spreading rate $\beta$. Results are obtained over $V = 5,000$ simulations. (d-f) Same as in (a), (b), and (c), respectively, but for $q=1,000$.}}
   \label{fig:heterodegree}
\end{figure*}

\change{In Fig.~\ref{fig:heterodegree}, we explore the effect of heterogeneous degree distributions in synthetic group-structured populations using edge-colored graphs with $N = 10,000$ nodes. We consider degrees following power-law degree distribution defined by $P(k) \sim k^{-\gamma}$ for $k \in [2, 10]$, and $P(k) = 0$ otherwise. We fix the mixing parameter $\mu = 0.025$ and community sizes at $q=100$. We vary the exponent values $\gamma$ from $2$ to $4$ for graphs with either maximally correlated (solid curves) or uncorrelated (dotted curves and transparent symbols) group structures. We consider three values of the spreading rate $\beta=0.2$, $0.4$, and $0.6$, corresponding to reproduction numbers $R_0=2.4$, $4.8$, and, $7.2$, respectively. We show (a) the outbreak size, (b) the peak infected population fraction, and (c) the total outbreak duration, all as functions of the exponent $\gamma$. These values are averages over $V = 5,000$ simulations. We see that the outbreak size and the peak infected population reduce as the degree exponent $\gamma$ increases, i.e., as the degree distribution becomes less heterogeneous.}

\begin{figure*}[!htb]
   \centering
   \includegraphics[width=0.85\textwidth]{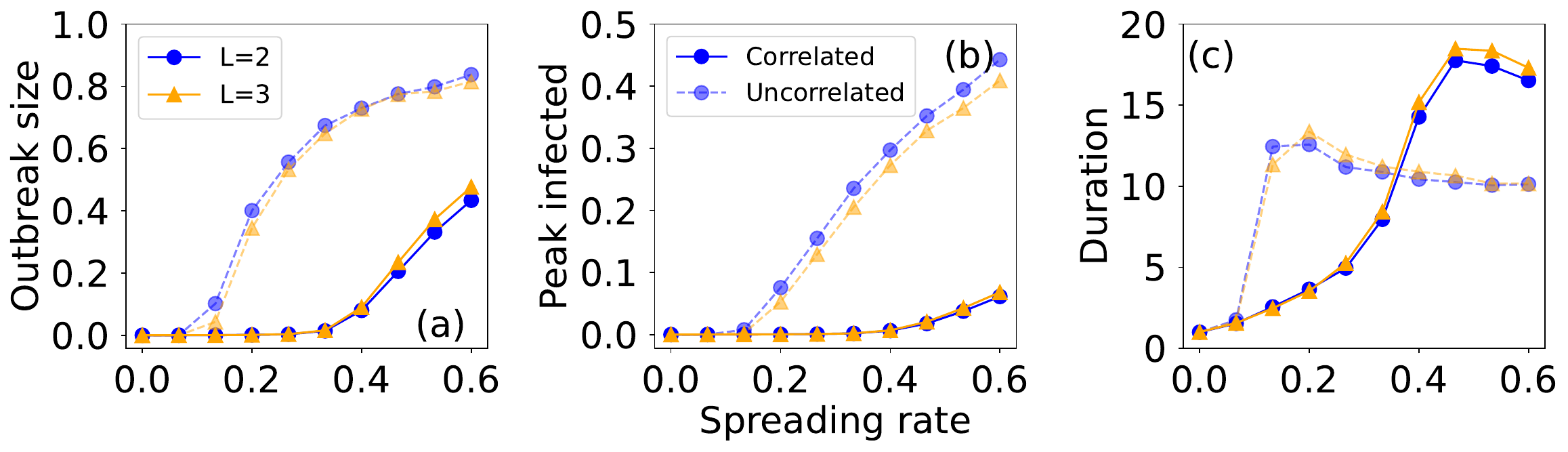}
   \caption{\change{\textbf{Epidemic spreading in synthetic group-structured populations.} We consider edge-colored graphs with $N = 10,000$ nodes. We consider two versions of the system, respectively composed of two and three layers. In both versions, the group sizes $q$ are set equal to $20$ in all the layers and the mixing parameter is $\mu = 0.025$. In the two-layer case, we fix $\langle k^{(1)} \rangle = \langle k^{(2)}\rangle = 6$, whereas in the three-layer case we have  $\langle k^{(1)} \rangle = \langle k^{(2)}\rangle = \langle k^{(3)}\rangle = 4$. We contrast the cases where the layers have a completely correlated and uncorrelated community structure. We tune the pairwise correlation among the community structure of the layers by swapping community memberships of nodes as explained in the Methods section. 
   We run $V = 5,000$ simulations of the SIR dynamics and display average values in the plot. (a) Outbreak size, (b) fraction of the infected population at the peak, and (c) total duration of the outbreak are plotted as functions of the spreading rate $\beta$.}}
   \label{fig:3layer}
\end{figure*}

\change{\subsection*{Edge-colored graphs with three layers}

We study spreading on three-layer edge-colored graphs with correlated and uncorrelated group structures. We consider networks with $N = 10,000$, the average degrees on the three layers is set to $\langle k^{(1)} \rangle = \langle k^{(2)}\rangle = \langle k^{(3)}\rangle = 4$, and the mixing parameter is set to $\mu = 0.025$. We fix the group sizes $q=20$ in all the layers and compare the case where the three layers have a completely correlated and uncorrelated community structure. We tune the pairwise correlation among the community structure of the layers by swapping community memberships of nodes as explained in the Methods section. We run $V = 5,000$ simulations of the SIR dynamics and display average values in Figure \ref{fig:3layer}. We further compare the results obtained on the three-layer edge-colored graphs with the two-layer case. To make the results comparable, we set $\langle k^{(1)} \rangle = \langle k^{(2)}\rangle = 6$, so that the overall average degree of the nodes is the same as for the three-layer edge-colored graphs.}

\subsection*{Site percolation on edge-colored graphs}

In the ordinary site-percolation model, 
nodes are randomly occupied with probability $p$. For $p=0$, no nodes are present in the system, so all clusters have null size; for $p=1$, only one single cluster, coinciding with the whole network, is present. The term percolation transition refers to the structural change, between these two extreme configurations, observed as a function of the occupation probability $p$. 
Please note that no actual phase transition occurs in a network of finite size. We can, however, still well describe the overall change in connectedness of the network by looking at how the
relative size of its largest cluster
varies with $p$, and identify the pseudo-critical point $p_c$
as value of the occupation probability $p$ where the ratio of the standard deviation and the average value of the largest cluster, across multiple realizations of the model, is maximal. 

To mimic
the effect of random immunization on edge-colored graphs, we compute the site-percolation threshold of the aggregated version of the graph, i.e., the single-layer network where nodes are connected if there exists an edge between them in any of the layers. 

\begin{figure}[!htb]
   \centering
   \includegraphics[width=0.4\textwidth]
   {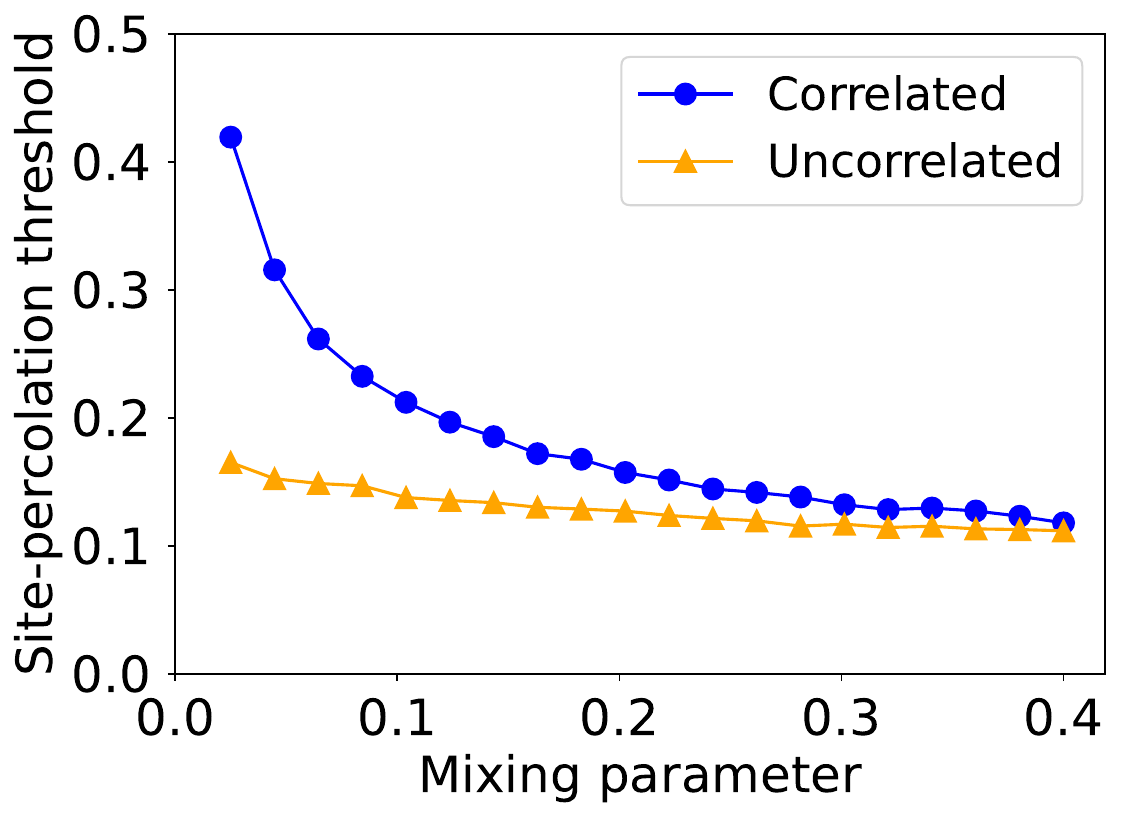}
   \caption{\textbf{Effect of community correlation on the site-percolation threshold of edge-colored graphs.} Edge-colored graphs with label shuffling probability $r=0$ (correlated community structure) and $r=1$ (uncorrelated community structure) are considered. The number of nodes $N=10,000$ nodes, average degrees $k^{(1)}=3$ and $k^{(2)}=10$, community sizes $q^{(1)}=5$ and $q^{(2)}=25$ in the two layers. The site percolation thresholds of the aggregated edge-colored graphs with correlated (blue circles) and uncorrelated (orange triangles) community structures are plotted as a function of the mixing parameter ($\mu$).}
   \label{fig:siteperc}
\end{figure}

We use the same experimental setting as in Figs.~2 (d-f) of the main paper: number of nodes $N=10,000$ nodes, average degrees $k^{(1)}=3$ and $k^{(2)}=10$, community sizes $q^{(1)}=5$ and $q^{(2)}=25$. We consider edge-colored graphs with label shuffling probabilities $r=0$ (correlated community structure) and $r=1$ (uncorrelated community structure). We plot the site-percolation thresholds of the aggregated edge-colored graphs with correlated (blue circles) and uncorrelated (orange triangles) community structures as a function of the mixing parameter ($\mu$) in Fig.~\ref{fig:siteperc}. We use the Newman-Ziff algorithm to 
efficiently simulate the site-percolation model~\cite{newman2001fast}.


\section*{Mean-field approximations}

We consider two mean-field approximations for continuous-time susceptible-infected-recover (SIR) dynamics~\cite{Pastor-Satorras2015} on an edge-colored graph composed of two layers~\cite{ramsey1987problem}. Each layer is generated according to the planted partition model (PPM)~\cite{condon1999algorithms}. Nodes are partitioned in $Q$ groups of identical size $q=N/Q$. We denote with $\sigma_i^{(\ell)} = 1, \ldots, Q$ the community membership of node $i$ in layer $\ell$. Two generic nodes $i$ and $j$ in  layer $\ell$ are connected with probability $p_{in} = \langle k \rangle (1-\mu)/ (q-1)$ if $\sigma_i^{(\ell)} = \sigma_j^{(\ell)}$, and with probability 
$p_{out} = \langle k \rangle \mu /(N-q)$ if $\sigma_i^{(\ell)} \neq \sigma_j^{(\ell)}$. The parameter $\langle k \rangle$ denotes the average number of connections of nodes in a layer; $\mu$ is the mixing parameter of the PPM. 

We initially assume that $\sigma_i^{(1)} = \sigma_i^{(2)}$ for all $i = 1, \ldots, N$. This ensures perfect correlation among the layer-wise partitions.
Correlation is reduced by shuffling, with probability $0 \leq r \leq 1$, the community membership of each node in layer $\ell=1$ with that of another randomly selected node. For $r=0$,  partitions are still perfectly correlated; for $r=1$, no correlation exists between the two partitions.

We denote with $\beta$ and $\gamma$ the spreading and recovery rates of SIR dynamics, respectively.

\subsection*{Group-based mean-field approximation (GBMFA)}

Since the system is perfectly symmetric, meaning that the number of communities in each layer equals $Q$, and the probability of observing respectively within-group and between-group connections are $p_{in}$ and $p_{out}$, then we can consider that there are only two types of nodes:
1) nodes with identical community labels and 2) nodes with different community labels. Let us indicate with $n_1$ the number of nodes in each of the 
groups with identical labels, and with $n_2$ the number of nodes in each of the groups with distinct labels. There are a total of $Q$ groups of nodes with identical labels, and $Q^2-Q$ groups of nodes with distinct group labels.

The expected number of nodes of type 1 is
\[
n_1 = \frac{(1-r) \, N}{Q} + \frac{r \, N}{Q^2} \; .
\]
Before nodes are re-assigned, there are $N/Q$ nodes with identical community labels $(a,a)$ for each $a$ value. Out of these nodes, $(1-r)N/Q$ are not re-assigned. Among those that are re-assigned, i.e., $r N /Q \times 1/Q$ are re-assigned to their original cluster, thus the second term. 
The expected number of nodes of type 2 is obtained from the constraint
$N = Q n_1 + Q(Q-1)n_2$, thus
\[
n_2 =\frac{N - Q n_1}{Q^2 - Q} \; .
\]

\subsubsection*{Connectivity matrix for type-1 and type-2 nodes.}

We first focus on the number of spreading events that, on average, a generic node of type 1 is involved with in the network. Indicate with $(a,a)$ the cluster labels of this generic node. We can identify four main types of spreading events that this node has with other nodes in the network:

\begin{enumerate}
\item A spreading event between a node of type 1 with community labels $(a,a)$ and another node of type 1 with community label $(a,a)$ happens at rate $2 \beta p_{in}$. This follows from the fact that 
the two nodes are connected with probability $p_{in}$ in each layer, and that there are two layers. There are $n_1-1$ of such potential spreading events as there are other $n_1-1$ nodes with community labels $(a,a)$.

\item A spreading event between a node of type 1 with cluster labels $(a,a)$ and another node of type 1 with community label $(b,b)$, with $a \neq b$, happens at rate $2 \beta p_{out}$. There are $(Q-1) n_1$ of such potential spreading events as there are other $Q-1$ pairs of community labels  $(b,b)$, with $a \neq b$, each containing $n_1$ nodes.

\item A spreading event between a node of type 1 with community labels $(a,a)$ and another node of type 2 with community labels $(a,b)$ or $(c,a)$, with $b \neq a$ and $c \neq a$, happens at rate $\beta (p_{in}+p_{out})$.
The term $p_{in}$ is due to the fact that the two nodes belong to the same community in one of the layers; 
the term $p_{out}$ is due to the fact that the two nodes belong to  distinct communities in one of the layers.
There are $2 (Q-1) n_2$ of such potential spreading events. There are in fact $Q-1$ groups with labels $(a,b)$ and $b \neq a$, and other $Q-1$ groups with labels $(c,a)$ and  $c \neq a$. Each of these groups contain  $n_2$ nodes on average.

\item Finally, a spreading event between a node of type 1 with community labels $(a,a)$ and another node of type 2 with community labels $(b,c)$, with $c \neq b$, $c \neq a$ and $b \neq c$, happens at rate $2 \beta p_{out}$ and there are $(Q^2-3Q+2) \, n_2$ potential events of this type. The coefficient $Q^2-3Q+2 = Q(Q-1) - 2 (Q-1)$, where $Q(Q-1)$ is the total number of pairs of distinct community labels, and we discounted from this number the other communities already considered at point 3.
\end{enumerate}

We summarize the various types of spreading events involving a node of type 1 in Table~\ref{tab:1}.

\begin{table}[!htb]
    \centering
    \begin{tabular}{c|c|c}
    Type of node & Number of contacts & Effective spreading rate\\
    \hline
    \hline
         $1, (a,a)$ & $n_1-1$ & $2 \beta p_{in}$  \\
         \hline
         $1, (b,b)$ & $(Q-1) n_1$ & $2 \beta p_{out}$  \\
         \hline
         $2, (a,b) \lor (b,a)$  & $2 (Q-1) n_2$ & $\beta (p_{in}+p_{out})$  \\
         \hline
         $2, (b,c)$ & $(Q^2-3Q+2) \, n_2$ & $2 \beta p_{out}$ 
    \end{tabular}
    \caption{Connectivity matrix for a node of type 1 with community labels $(a,a)$.}
    \label{tab:1}
\end{table}

We can now repeat a similar enumeration for a node of type 2 with community labels   $(a,b)$ and $a \neq b$.  We can identify five main types of spreading events that this node has with other nodes in the network:

\begin{enumerate}
\item A spreading event between a node of type 2 with community labels $(a,b)$, with $a \neq b$, and another node of type 1 with community labels $(a,a)$ or $(b,b)$ happens at rate $\beta (p_{in} + p_{out})$. The term $p_{in}$ is due to the fact that the two nodes belong to the same community in one of the layers; 
the term $p_{out}$ is due to the fact that the two nodes belong to  distinct communities in one of the layers. There are $2 n_1$ of such potential spreading events as there are only $2$ groups of this type each composed of $n_1$ nodes.

\item A spreading event between a node of type 2 with community labels $(a,b)$, with $a \neq b$, and another node of type 1 with community labels $(c,c)$, with $a \neq c$ and $b \neq c$, happens at rate $2 \beta p_{out}$.
This follows from the fact that 
the two nodes are connected with probability $p_{out}$ in each of the two layers. There are $(Q-2) n_1$ of such potential spreading events as there are only $Q-2$ groups of this type each composed of $n_1$ nodes.

\item A spreading event between a node of type 2 with community labels $(a,b)$, with $a \neq b$, and another node of type 2 with community labels $(a,b)$ happens at rate $2 \beta p_{in}$.
This follows from the fact that 
the two nodes are connected with probability $p_{in}$ in each of the two layers. There are $n_2$ of such potential spreading events as there is only $1$ group of this type each composed of $n_2$ nodes.

\item A spreading event between a node of type 2 with community labels $(a,b)$, with $a \neq b$, and another node of type 2 with community labels $(a,c)$ or $(c,b)$, with $a \neq c$ and $b \neq c$, happens at rate $ \beta (p_{in} + p_{out})$. This follows from the fact that 
the two nodes are connected with probability $p_{in}$ in one layer, and with probability $p_{out}$
in the other layer. There are are $2 (Q-2) n_2$ of such potential spreading events as there are only $2(Q-2)$ groups of this type each composed of $n_2$ nodes.

\item Finally, a spreading event between a node of type 2 with community labels $(a,b)$, with $a \neq b$, and another node of type 2 with community labels $(c,d)$, with $a \neq c$, $b \neq d$ and $c \neq d$, happens at rate $ \beta p_{out} $. This follows from the fact that 
the two nodes are connected with probability $p_{out}$ in each of the two layers.
There are $(Q^2 - 3CQ + 3) \, n_2$ potential events of this type. The coefficient $Q^2 - 3Q + 3 = Q(Q-1) - 2 (Q-2) -1 $, where $Q(Q-1)$ is the total number of pairs of distinct community labels, and we discounted from this number the other clusters already considered at points 3 and 4.
\end{enumerate}

We summarize the various types of spreading events involving a node of type 2 in Table~\ref{tab:2}.

\begin{table}[]
    \centering
    \begin{tabular}{c|c|c}
    Type of node & Number of contacts & Effective spreading rate\\
    \hline
    \hline
         $1, (a,a) \lor (b,b)$ & $2 \, n_1$ & $\beta (p_{in}+p_{out})$  \\
         \hline
         $1, (c,c)$ & $(Q-2) n_1$ & $2 \beta p_{out}$  \\
         \hline
         $2, (a,b)$  & $n_2$ & $2 \beta p_{in}$  \\
         \hline
         $2, (a,d) \lor (c, b)$ & $2 (Q-2) \, n_2$ & $\beta (p_{in} + p_{out})$ 
         \\
         \hline
         $2, (c,d)$ & $ (Q^2-3Q+3) \, n_2$ & $2 \beta p_{out}$ 
    \end{tabular}
    \caption{Connectivity matrix for a node of type 2 with community labels $(a,b)$ with $a \neq b$.}
    \label{tab:2}
\end{table}

\subsubsection*{Mean-field equations}

The differential equation that determines the evolution of individual nodes of type 1 and 2 are respectively
\[
\frac{dS_1}{dt} = - \beta \, S_1 \,  [ 2 p_{in} S_1 (S_1 -1) + 2 p_{out} (Q-1) I_1 n_1  + 2 (C-1) I_2 n_2  (p_{in}+p_{out}) +  2 (Q^2-3Q+2) I_2 n_2 p_{out}]  
\]
and
\[
\begin{array}{ll}
\frac{dS_2}{dt} = & - \beta \, S_2 \,  [ 2 I_1 n_1  (p_{in}+p_{out}) + 
2 p_{out} (Q-2) I_1 n_1  
\\
& + 2  I_2 n_2 p_{in} + 2 (Q-2) I_2 n_2 (p_{in}+p_{out}) + 2 (Q^2-3Q+3) I_2 n_2 p_{out} 
]  \; ,
\end{array}
\]
where $S_x$ and $I_x$ are the probabilities that a generic node of type $x=1,2$ is susceptible or infected, respectively. The product $I_x n_x$ indicates the expected number of infected nodes in a group of nodes of type $x$ .
If we assume that $n_1 \gg 1$, re-arrange the terms on the rhs of the above equations, and perform some simple calculations, we
can simply write the common equation
\begin{equation}
\frac{dS_x}{dt} = - 2 \beta \, S_x \,  \left[ I_1 n_1 + I_2 n_2 (Q-1) \right] \left[ p_{in} + (Q-1) p_{out} \right]
\label{eq:gbmf1}
\end{equation}
with $x = 1, 2$.
Also, we have
\begin{equation}
\frac{dI_x}{dt} = - \frac{dS_x}{dt} - \gamma I_x \; ,
\label{eq:gbmf2}
\end{equation}
with $x = 1, 2$. 
Assuming that the initial condition of the dynamics is such that $I_x(0) = I(0)$ and $S_x(0) = S(0)$ for  $x = 1, 2$, then we 
have $S_x(t) = S(t)$ and $I_x(t) = I(t)$ for both $x=1, 2$, thus we can  write
\[
\frac{dS}{dt} = - 2 \beta \, S \, I \, \left[  n_1 +  n_2 (Q-1) \right] \left[ p_{in} + (Q-1) p_{out} \right]
\]
and
\[
\frac{dI}{dt} = - \frac{dS}{dt} - \gamma I \; .
\]

If we assume that $S \simeq 1$ and $I \ll 1$, 
we can linearize the above equations to determine the
epidemic threshold as
\[
[\beta/\gamma]_c = \left\{2 \, \left[ n_1 + n_2 (Q-1) \right] \left[ p_{in} + (Q-1) p_{out} \right] \right\}^{-1}   \; .
\]
We remind that $n_1 + n_2 (Q-1) = N/Q$, and $N/Q \left[ p_{in} + (Q-1) p_{out} \right]  = \langle k \rangle$
is the average degree of a node in one of the layers. We thus end up with the solution
\begin{equation}
[\beta/\gamma]_c = \frac{1}{2 \langle k \rangle}  \; ,
\label{eq:critic_gb}
\end{equation}
stating that the value of the epidemic threshold doesn't depend on the specific choice of the parameter $r$, and it's simply equivalent to the one of an Erd\H{o}s-R\'enyi graph with average degree $2 \langle k \rangle$. 

\change{
The above derivation can be adapted to the case in which the layers are constructed with different values of the parameters $p^{(1)}_{in}$, $p^{(1)}_{out}$, $p^{(2)}_{in}$, and $p^{(2)}_{out}$. The adaptation is straightforward in the sense that we can re-use all the above equations by simply implementing the transformations: $2 p_{in} \to p^{(1)}_{in} + p^{(2)}_{in}$ and $2 p_{out} \to p^{(1)}_{out} + p^{(2)}_{out}$. Also in this case, we can write $N/Q \left[ p^{(\ell)}_{in} + (Q-1) p^{(\ell)}_{out} \right]  = \langle k^{(\ell)} \rangle$ for $\ell=1, 2$, thus the resulting epidemic threshold can be written as
\begin{equation}
[\beta/\gamma]_c = \frac{1}{\langle k^{(1)} \rangle + \langle k^{(2)} \rangle}  \; ,
\label{eq:critic_gb_diff}
\end{equation}
meaning that the value of the epidemic threshold doesn't depend on the specific choice of the parameter $r$, and it's simply equivalent to the one of an Erd\H{o}s-R\'enyi graph with average degree $\langle k^{(1)} \rangle + \langle k^{(2)} \rangle$. 
}

\subsection*{Individual-based mean-field approximation (IBMFA)}

This is the classical approach described in several papers, including the review paper by Pastor-Satorras {\it et al.}~\cite{Pastor-Satorras2015}.
We indicate with $A^{(1)}$ and $A^{(2)}$ the adjacency matrices of the two layers of interactions. Specifically, $A^{(\ell)}_{ij} = A^{(\ell)}_{ji} = 1$ is nodes $i$ and $j$ are connected with an edge in layer $\ell$, and $A^{(\ell)}_{ij} = A^{(\ell)}_{ji} =0$ otherwise. 
We describe the average dynamical behaviour of each node $i$ with the real-valued variables $S_i(t)$, $I_i(t)$ and $R_i(t)$ representing respectively the probability of finding node $i$ in the state S, I or R at time $t$. We clearly have $S_i(t) + I_i(t) + R_i(t) = 1$. The differential equations that describe the dynamics of the system are:
\begin{equation}
\begin{array}{l}
\frac{dS_i}{dt} = - \beta \, S_i(t) \, \sum_j \, \left(  A^{(1)}_{ji} + A^{(2)}_{ji} \right) \, I_j(t) 
\\
\frac{dI_i}{dt} =  \beta \, S_i(t) \, \sum_j \, \left(  A^{(1)}_{ji} + A^{(2)}_{ji} \right) \, I_j(t)  - \gamma I_i(t)
\\
\frac{dR_i}{dt} = \gamma I_i(t)
\end{array}
    \label{eq:imbfa}
\end{equation}

The fraction of nodes in the susceptible state is obtained as $S = \sum_i S_i / N$. Similar expressions are valid for the fraction  of infected and recovered nodes.

The epidemic threshold is estimated by performing a linear stability analysis of the above system of differential equations, from which one finds that
\begin{equation}
\left[\beta/\gamma\right]_c = \frac{1}{\lambda_{\max}} \;,
    \label{eq:imbfa_critical}
\end{equation}
where $\lambda_{\max}$ is the largest eigenvalue of the matrix $A = A^{(1)} + A^{(2)}$ corresponding to the aggregation of two network layers.

\begin{figure*}[!htb]
   \centering
   \includegraphics[width=0.85\textwidth]{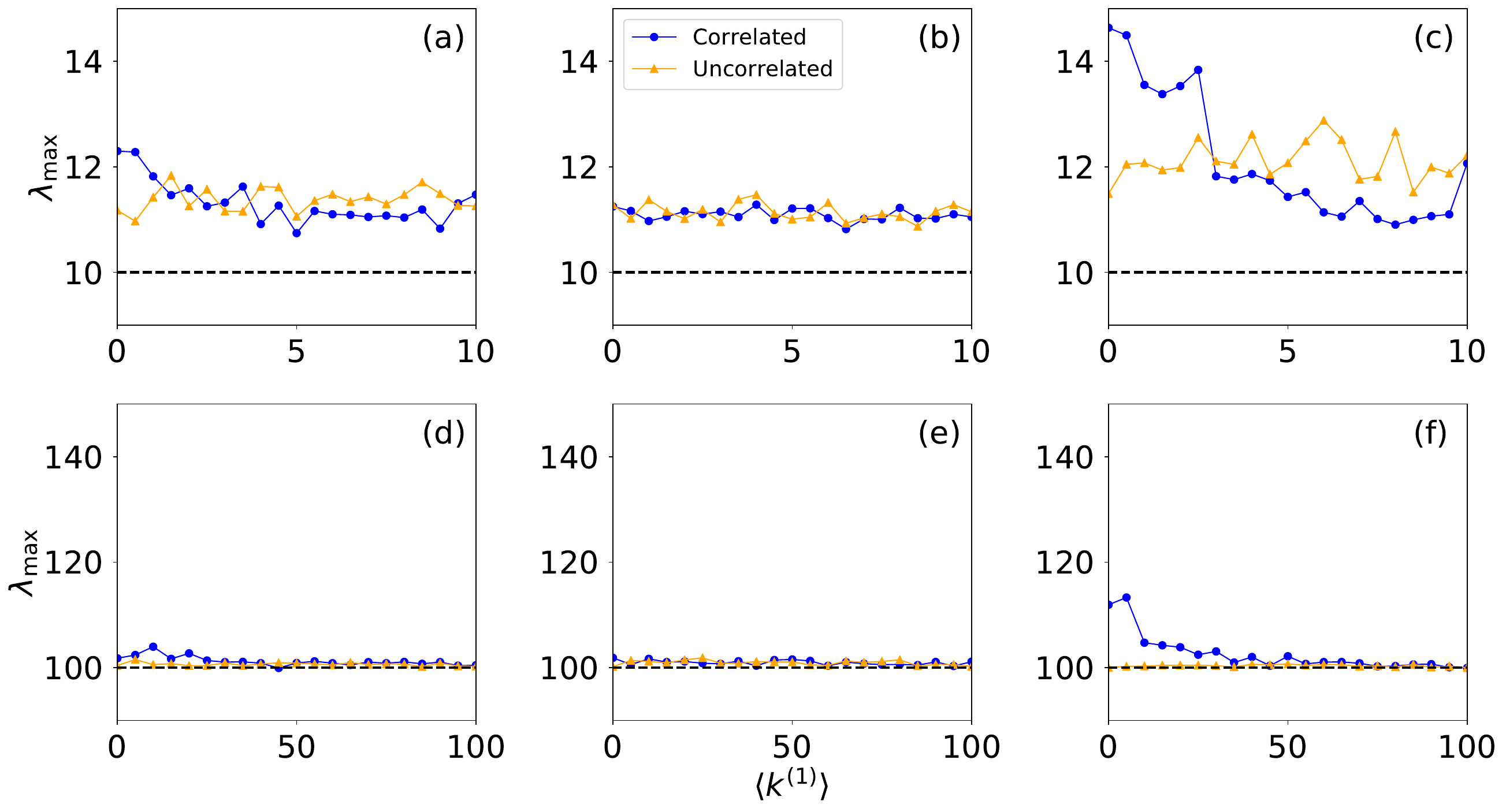}
   \caption{\newchange{\textbf{Theoretical prediction of the epidemic threshold in group-structure populations.}
   (a) We consider edge-colored graphs with $N=1,000$ nodes and label shuffling probabilities $r=0$ (correlated community structure, blue circles) and $r=1$ (uncorrelated community structure, orange triangles). We set the community sizes $q^{(1)} = q^{(2)} = q = 100$, the mixing parameter to $\mu = 0.025$ and the overall average degree $\langle k \rangle = 10$. One layer has average degree $\langle k^{(1)} \rangle$, whereas the other has average degree $\langle k^{(2)} \rangle = \langle k \rangle - \langle k^{(1)} \rangle$.  Once the graph has been generated, we numerically estimate the largest eigenvalue of the adjacency matrix $\lambda_{\max}$. We plot $\lambda_{\max}$ vs. $\langle k^{(1)} \rangle$. The dashed horizontal line indicates $\lambda_{\max} = \langle k \rangle$. (b) Same as in (a), but for $\mu = 0.3$. (c) Same as in (a), but for $q=25$.
   (d) Same as in (a), but for $\langle k \rangle = 100$ and $q=250$. (e) Same as in (d), but for $\mu = 0.3$. (f) Same as in (d), but for $q = 100$.
   }}
   \label{fig:eigs}
\end{figure*}

\change{If we assume that the network layers are instances of the Erd\H{o}s-R\'enyi model with average degree respectively $\langle k^{(1)} \rangle$ and $\langle k^{(2)} \rangle$, and that both  $\langle k^{(1)} \rangle$ and $\langle k^{(2)} \rangle$ are much smaller than the size 
\newchange{of the individual groups} so that the layers have negligible edge overlap, we can write $\lambda_{\max} = \langle k^{(1)} \rangle + \langle k^{(2)} \rangle$, from which we recover the same expression as in Eq.~(\ref{eq:critic_gb_diff}).}
\newchange{We verify the above prediction in numerical experiments, see Figure~\ref{fig:eigs}. Small deviations from our prediction are visible only for fully correlated group structure if $\langle k^{(1)} \rangle + \langle k^{(2)} \rangle$
is comparable with the size of the individual communities.}

\subsection*{Numerical validation of the mean-field approximations}

\begin{figure*}
    \centering
 \includegraphics[width=\textwidth]{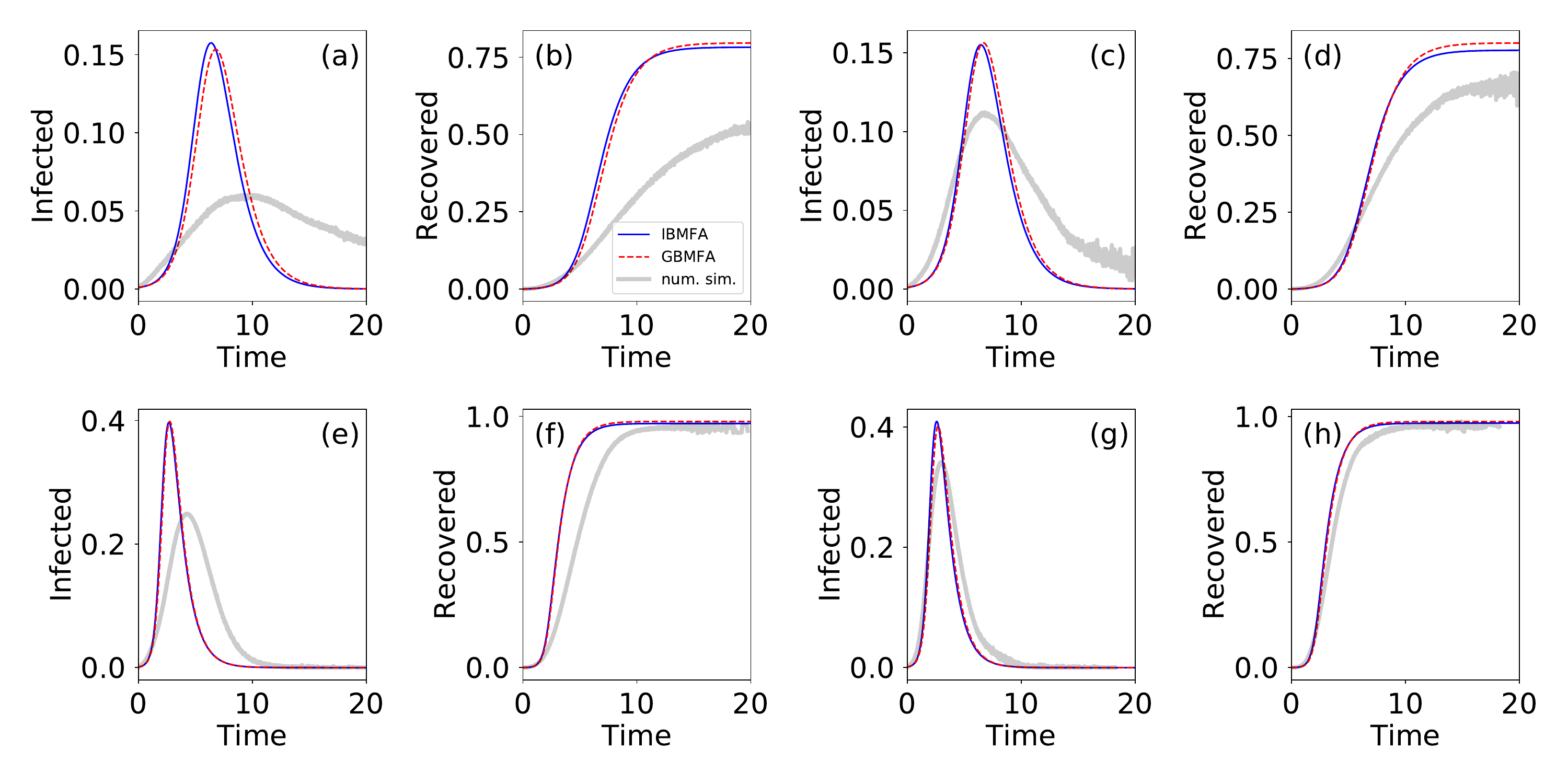}
    \caption{
    {\bf Comparison between mean-field predictions and numerical estimates of the epidemic curves.}
    (a) We plot the fraction of infected nodes as a function of time predicted by the group-based mean-field approximation (GBMFA), the individual-based mean-field approximation (IBMFA), and estimated from numerical simulations (averages only on surviving runs over a total of $V=5,000$ repetitions). Parameters of the model are $N=1,000$, $Q=10$, $\langle k \rangle = 10$, $r=0$ and $\beta =0.1$. (b) Same as in (a), but for the fraction of recovered nodes as a function of time. (c) Same as in (a), but for $r=1$.  (d) Same as in (b), but for $r=1$. (d-f) Same as in (a-d), respectively, but for $\beta=0.2$.
}
\label{fig:pred1}
\end{figure*}

\begin{figure*}[!htb]
    \centering    \includegraphics[width=\textwidth]{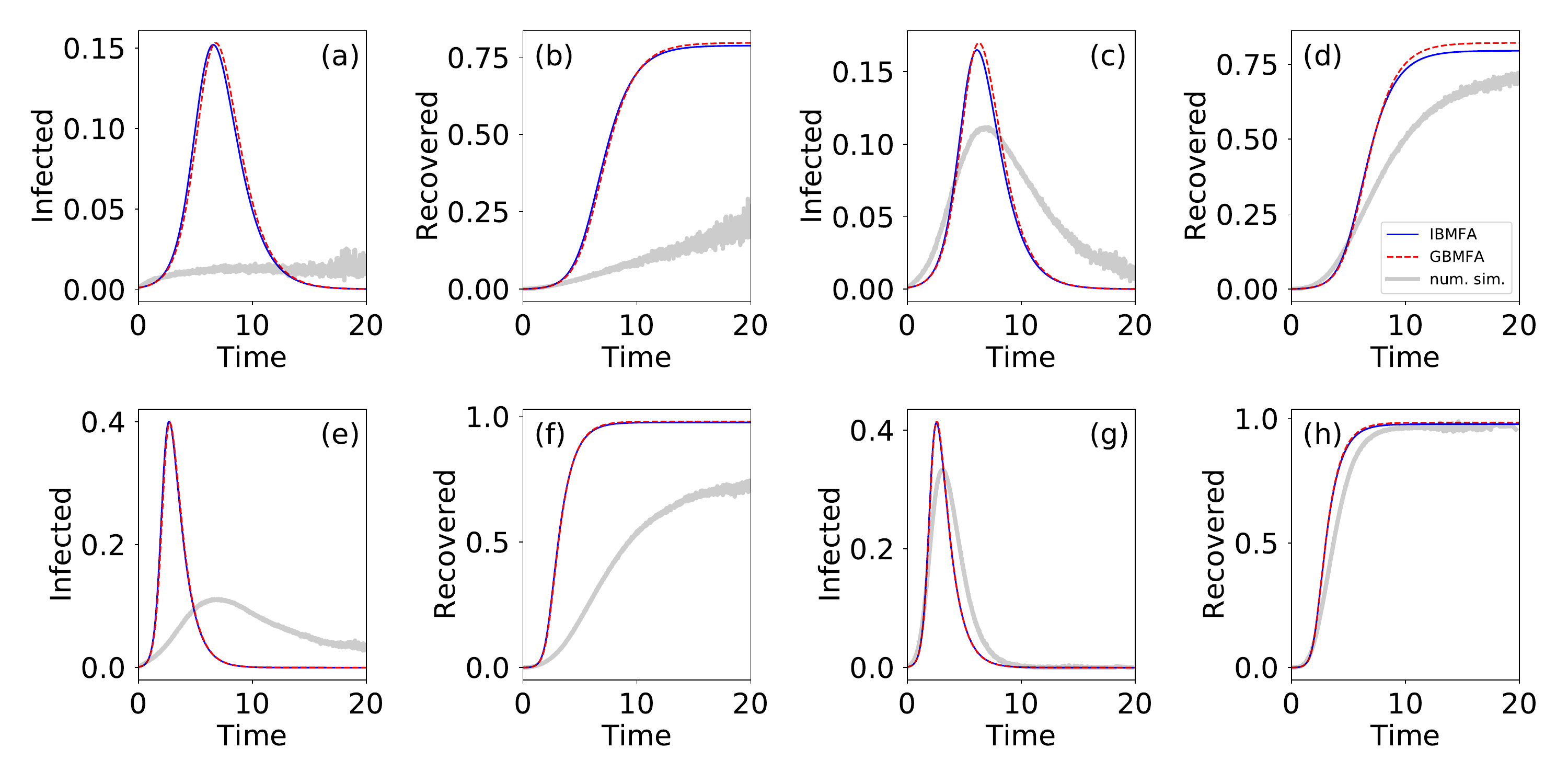}
    \caption{
    {\bf Comparison between mean-field predictions and numerical estimates of the epidemic curves.} Same as in Fig.~\ref{fig:pred1} but for $Q=50$.
}
\label{fig:pred2}
\end{figure*}

\begin{figure*}[!htb]
    \centering
    \includegraphics[width=\textwidth]{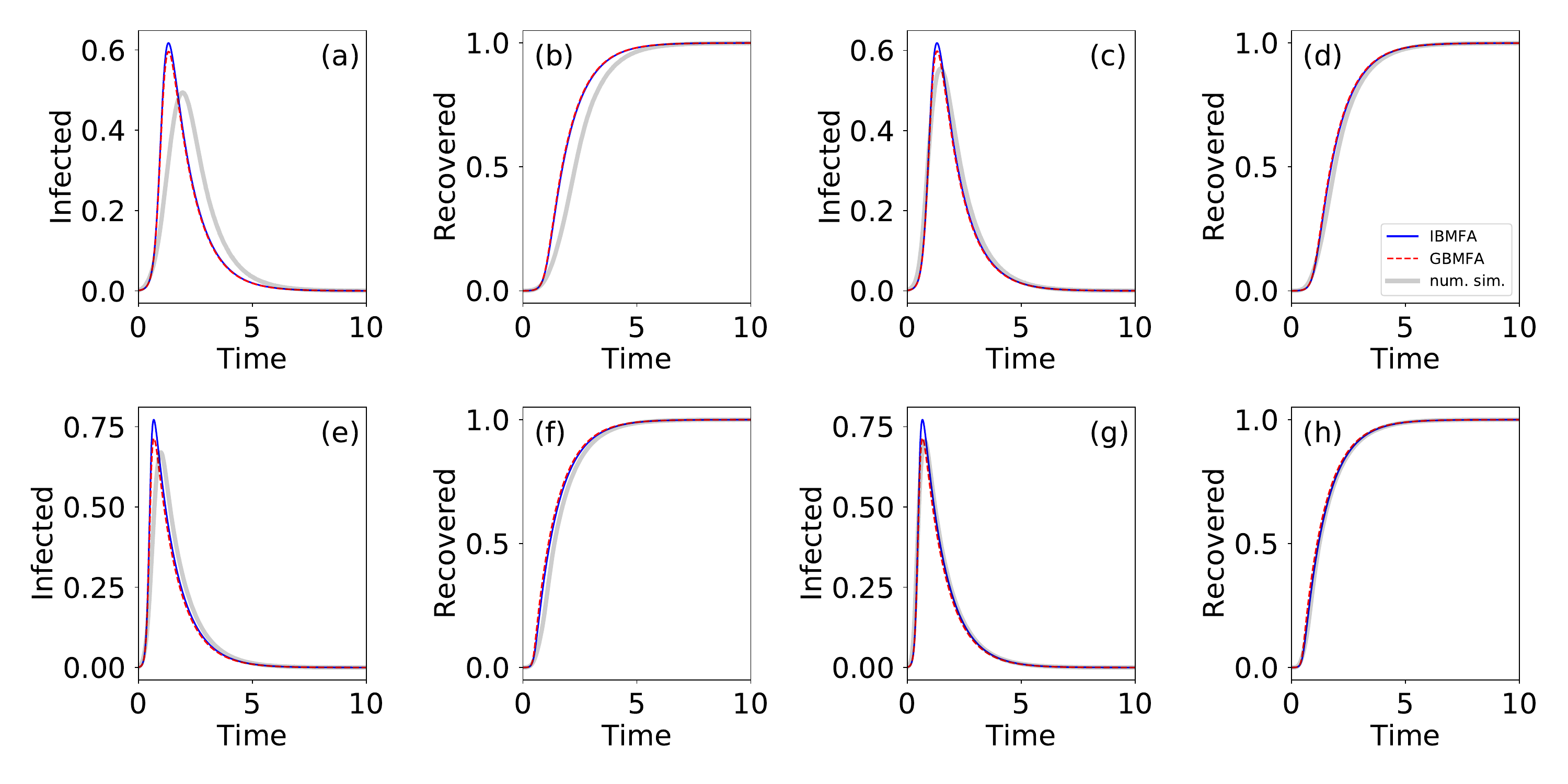}
    \caption{{\bf Comparison between mean-field predictions and numerical estimates of the epidemic curves.} Same as in Fig.~\ref{fig:pred1}, but for $\langle k \rangle = 40$.
}
\label{fig:pred3}
\end{figure*}

In Figs.~\ref{fig:pred1},~\ref{fig:pred2} and ~\ref{fig:pred3}, we compare GBMFA and IBMFA predictions against results of numerical simulations. Mean-field predictions are obtained from the numerical integration of Eqs.~\ref{eq:gbmf1} and~\ref{eq:gbmf2} for GBMFA, and of Eqs.~\ref{eq:imbfa} for IBMFA. Results of numerical simulations are displayed as average values within binned time intervals of length $\delta t = 10/N$; only surviving runs (i.e., realizations of the spreading process that contain at least one infected node at a given point in time) are used to compute the average within the bins. The large-time regime of the epidemic curves obtained from numerical simulations is due to the increasingly smaller number of surviving runs that are obtained as time increases. A total of $V=5,000$ simulations are run for each configuration. In all cases, SIR dynamics is started from an initial conditions in which all nodes are in the S state, except for a randomly chosen node in the I state.

\begin{figure*}[!htb]
    \centering
    \includegraphics[width=0.85\textwidth]{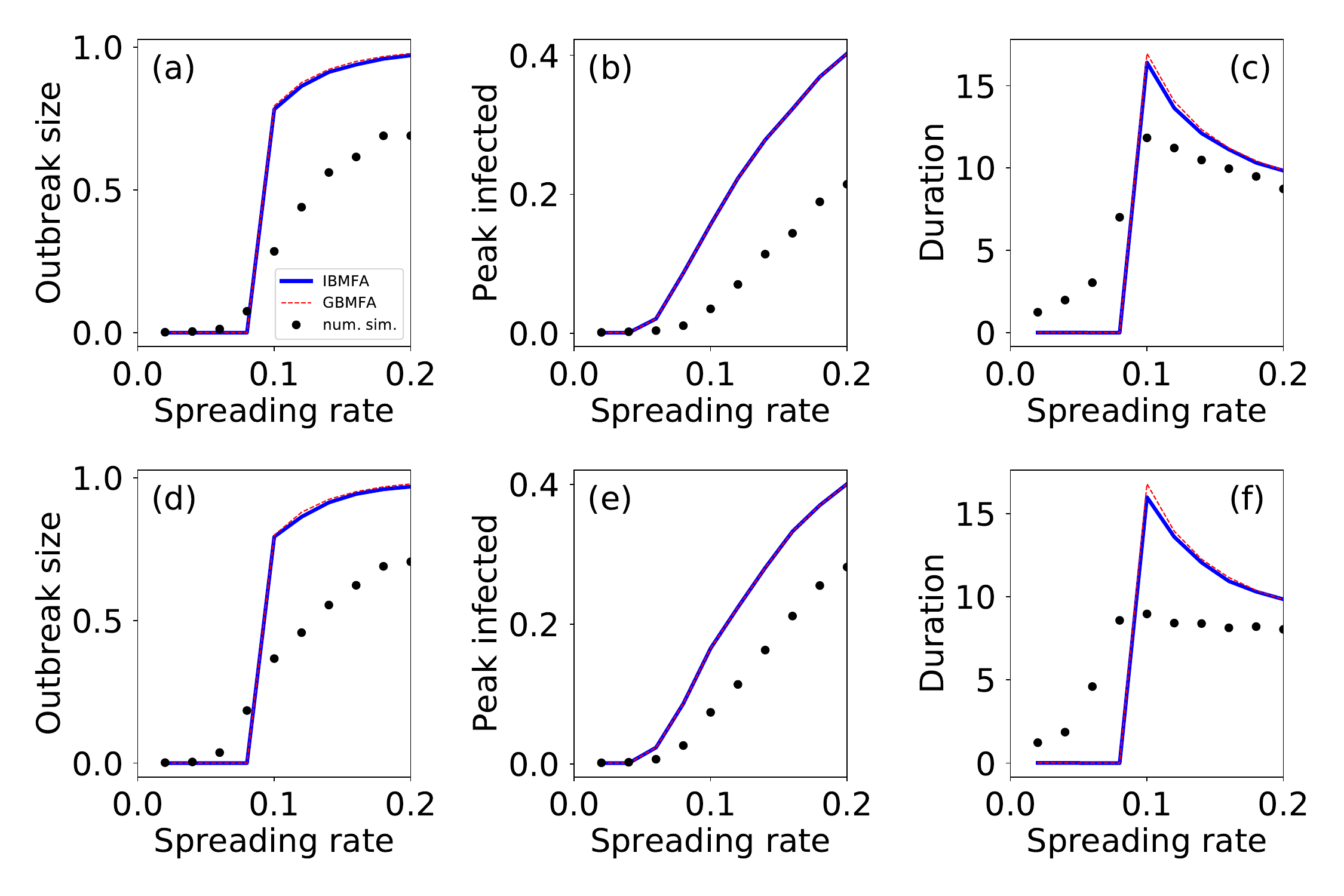}
    \caption{{\bf Comparison between mean-field predictions and numerical estimates of the metrics of epidemic severity.} (a) We consider the same setting as in Fig.~\ref{fig:pred1} and compute the size of the outbreak size for a given value of the spreading rate using $V = 1,000$ simulations of SIR dynamics. At the same time, we provide estimates of the outbreak size using also IBMFA and  GBMFA. (b) Same as in (a), but for the peak value of the infected. (c) Same as in (a), but for the duration of the spreading process. (d-e) Same as in (a-c), respectively, but for $r=1$.
}
\label{fig:pred4}
\end{figure*}

In Fig.~\ref{fig:pred4}, we compare mean-field predictions of the size of outbreak, the peak value of infected, and the duration of the epidemic with estimates from numerical simulations (averages over $V=1, 000$ realizations). Please note that the average of the peak value of the fraction of infected does not correspond to those visualized in Fig.~\ref{fig:pred1}. In the measurements done in Fig.~\ref{fig:pred4}, the peak value can occur at any instant of time during SIR dynamics, instead the results of Fig.~\ref{fig:pred1} display averages at specific ranges of time.

\subsection*{Group sizes for the IUB data}

\begin{figure*}[!htb]
    \centering
    \includegraphics[width=0.85\textwidth]
    {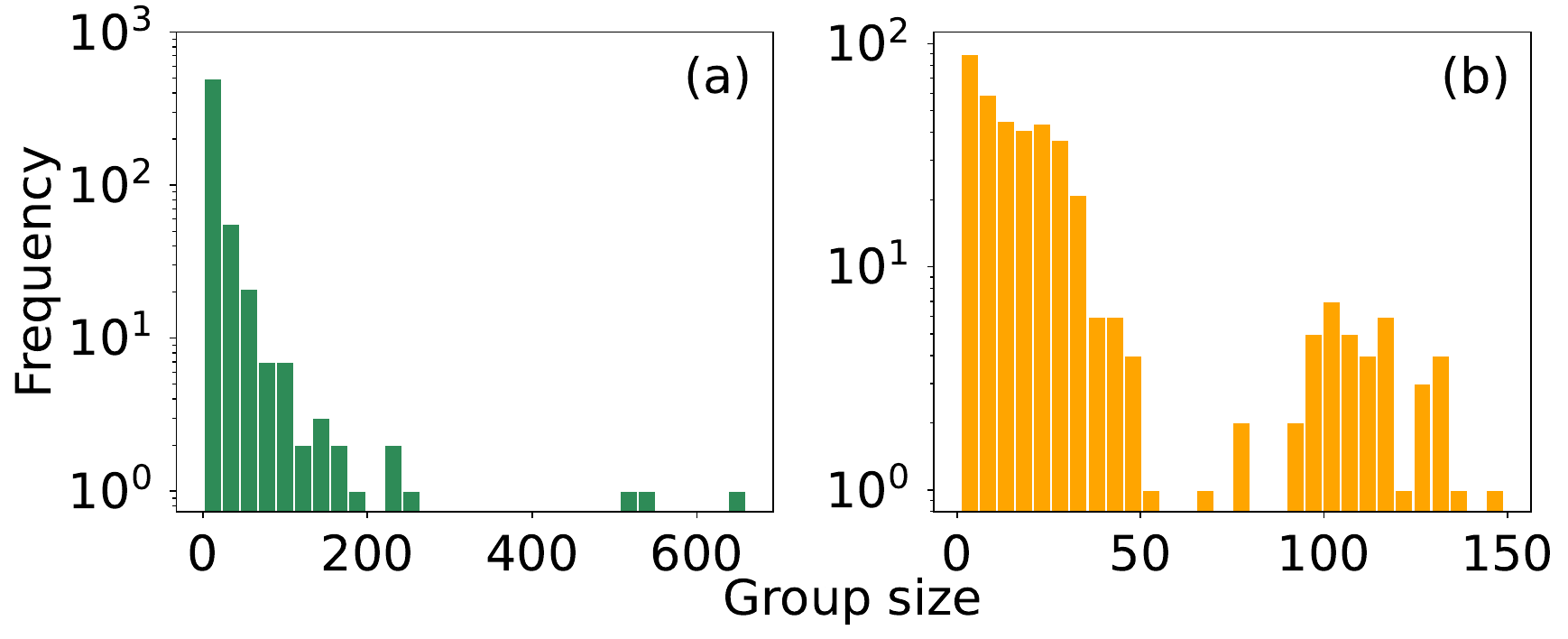}
    \caption{{\bf Frequency of group sizes for the IUB data.} (a) Frequency of the group sizes found in the Indiana University Bloomington (IUB) program/level data. (b) Frequency of the group sizes in the IUB housing data.}
\label{fig:iub_dist}
\end{figure*}

We take advantage of data about housing and attendance at Indiana University Bloomington (IUB) to construct realistic edge-colored group-structured graphs. 
There are $N=10,132$ individuals in the IUB dataset, each representing a student who resided in one of the campus facilities during the Fall 2019 semester. We use layer $\ell=1$ to represent contacts between students in housing settings. We form $Q^{(1)} = 396$ groups composed of students living on the same floor of a large dormitory, or in the same Greek house. The average size of these groups is $\langle q^{(1)} \rangle = 25.58$. Layer $\ell =2$ is used to represent classroom interactions. We form $Q^{(2)} = 600$ groups of students based on their enrollment program (e.g., Computer Science, Finance, Mathematics, Physics) and their education level (from Freshman to Ph.D.). The average size of these groups is $\langle q^{(2)} \rangle = 16.86$. The frequencies of the obtained group sizes for attendance and housing are shown in Figs.~\ref{fig:iub_dist} (a) and (b), respectively. 

\begin{figure*}[!htb]
    \centering
    \includegraphics[width=0.85\textwidth]{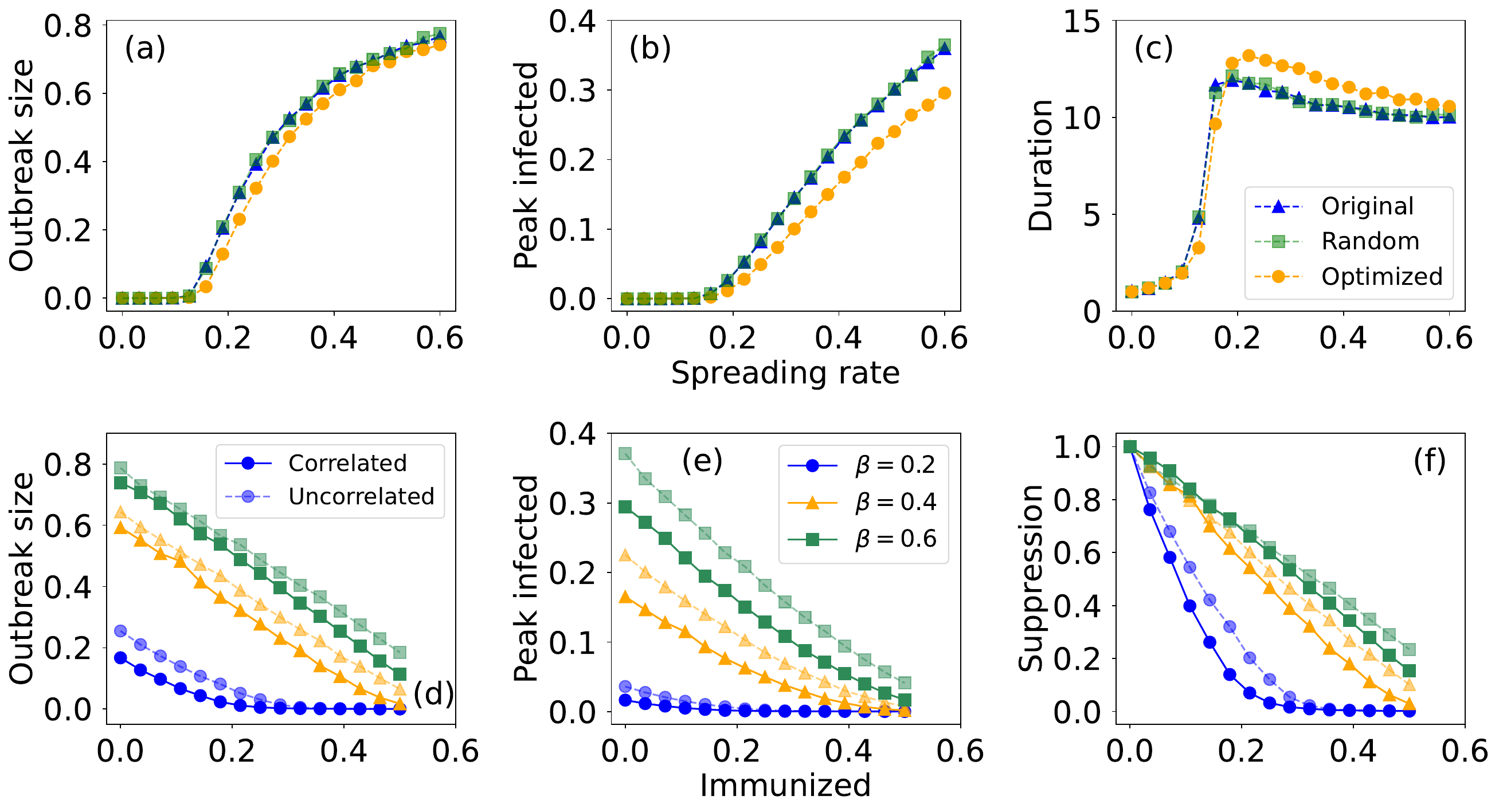}
    \caption{\change{{\bf Epidemic spreading in the student population of the Indiana University Bloomington.}
    (a) We use data about housing and attendance for the Fall 2019 semester at the Indiana University Bloomington (IUB) campus to generate edge-colored graphs with block structure. The community partition in one layer reflects housing assignments; the partition in the other layer serves to group students based on their program and education level.  Different graphs are generated depending on whether network partitions are (i) those directly observed from the data, (ii) randomized, or (iii) optimized for maximum correlation. We then simulate SIR dynamics on the graphs and measure the average size of the outbreak as a function of the spreading rate $\beta$. Results are averaged over $V=5,000$ repetitions. (b) Same as in (a), but for the peak fraction of infected. (c) Same as in (a), but for the duration of the spreading process. (d) We plot the size of the outbreak as a function of the fraction of individuals that are initially immunized. \change{We consider three values of the spreading rate $\beta$, i.e., $0.2$, $0.4$, and $0.6$, corresponding to the reproduction number $R_0$ equal to $2$, $4$, and $6$.} Different symbols correspond to different $\beta$ values; full curves and solid symbols indicate the optimized configuration considered in panel (a); dashed curves and transparent symbols refer to graphs created using ground-truth partitions.  Results are averaged over $V=5,000$ repetitions. (e) Same as in (d), but for the peak fraction of infected. (f) Same as in (d), but with abscissa values rescaled by the outbreak size observed when zero individuals are initially immunized.
}}
    \label{fig:real_0.1}
\end{figure*}

\change{\subsection*{College housing/attendance network}

In Fig.~\ref{fig:real_0.1}, we display results for the same experimental setting as in Fig.~4 of the main paper but for $\mu=0.1$. 

}

\end{document}